\documentclass[aps,twocolumn,prb,longbibliography,floatfix,showpacs,citeautoscript]{revtex4-1}
\usepackage{amsmath}
\usepackage{amssymb}
\usepackage{graphicx}
\usepackage{epsf}
\usepackage{epsfig}
\usepackage{braket}
\usepackage{bm}
\usepackage{amsfonts}
\usepackage{color,soul}
\usepackage{txfonts}

\newcommand{\ve}{\varepsilon}

\newcommand{\Ef}{E_{\text{F}}}

\newcommand{\bea}{\begin{eqnarray}}
\newcommand{\eea}{\end{eqnarray}}
\newcommand{\Ha}{\mathcal{H}}
\newcommand{\up}{\uparrow}
\newcommand{\dn}{\downarrow}

\begin{document}

\title{Gate induced enhancement of spin-orbit coupling in dilute fluorinated graphene}

\author{R. M. Guzm\'an-Arellano}
\affiliation{Centro At{\'{o}}mico Bariloche and Instituto Balseiro, CNEA, 8400 Bariloche, Argentina}
\affiliation{Consejo Nacional de Investigaciones Cient\'{\i}ficas y T\'ecnicas (CONICET), Argentina}
\author{A. D. Hern\'andez-Nieves}
\affiliation{Centro At{\'{o}}mico Bariloche and Instituto Balseiro, CNEA, 8400 Bariloche, Argentina}
\affiliation{Consejo Nacional de Investigaciones Cient\'{\i}ficas y T\'ecnicas (CONICET), Argentina}
\author{C. A. Balseiro}
\affiliation{Centro At{\'{o}}mico Bariloche and Instituto Balseiro, CNEA, 8400 Bariloche, Argentina}
\affiliation{Consejo Nacional de Investigaciones Cient\'{\i}ficas y T\'ecnicas (CONICET), Argentina}
\author{Gonzalo Usaj}
\affiliation{Centro At{\'{o}}mico Bariloche and Instituto Balseiro, CNEA, 8400 Bariloche, Argentina}
\affiliation{Consejo Nacional de Investigaciones Cient\'{\i}ficas y T\'ecnicas (CONICET), Argentina}
\date{\today}

\begin{abstract}
We analyze the origin of spin-orbit coupling (SOC) in fluorinated graphene using Density Functional Theory (DFT) and a tight-binding model for the relevant orbitals. As it turns out,  the dominant source of SOC is the atomic spin-orbit of fluorine adatoms and not the impurity induced SOC based on the distortion of the graphene plane as in hydrogenated graphene. More interestingly, our DFT calculations  show that SOC is strongly affected by both the type and concentrations of the graphene's carriers, being enhanced by electron doping  and reduced by hole doping. This effect is due to the charge transfer to the fluorine adatom and the consequent change in the fluorine-carbon bonding. Our simple tight-binding model, that includes the SOC of the $2p$ orbitals of F and effective parameters based on maximally localized Wannier functions, is able to account for the effect. The strong enhancement of the SOC induced by graphene doping opens the possibility to tune the spin relaxation in this material.
\end{abstract}




\maketitle

\section{Introduction}
Spin based transport phenomena in graphene is a flourishing area of research due to the expected long spin relaxation lengths resulting from the small spin-orbit coupling (SOC) of carbon atoms and the high carriers mobility. Due to these exceptional properties, the great potential of graphene for the study of new fundamental phenomena and for applications in nanoelectronics are now extended to the study of spin dependent phenomena with possible applications on spintronics.\cite{CastroNeto2009b,DasSarma2011,Zutic2004,Roche2014}  In this way the two-dimensional material, with unusual electronic properties,  became a model system in which the big challenge is to control and manipulate both, the charge and the spin degrees of freedom.     
Concerning the spin properties, the actual value of the spin relaxation rate of graphene carriers is still a controversial issue: while the spin relaxation times were initially thought to be very long,\cite{Huertas-Hernando2006,Min2006,Huertas-Hernando2009,Ertler2009,CastroNeto2009a} different experiments\cite{Tombros2007,Han2010a,Han2011,Zomer2012,Maassen2012,Lundeberg2013} suggest that they are much shorter than the theoretical predictions. This apparent controversy has recently been challenged by the way experiments have been interpreted.\cite{Idzuchi2014}

Whatever the source of the spin relaxation mechanism, an important issue for spintronics applications is the ability to control it.\cite{Zutic2004,Roche2014} Sources of spin relaxation in graphene are usually attributed to magnetic defects (vacancies\cite{Yazyev2007,Palacios2008,Yazyev2008,Balog2009,Yazyev2010,Haase2011} or adatoms\cite{Uchoa2008,Lehtinen2003,Duplock2004,Meyer2008,Chan2008,Uchoa2008,CastroNeto2009c,Boukhvalov2008,Boukhvalov2009,CastroNeto2009a,Cornaglia2009,Wehling2009,Wehling2010,Wehling2010a,Ao2010,Chan2011,Sofo2012}), to the presence of extrinsic (adatom induced) SOC\cite{CastroNeto2009a,Kochan2014} ripples\cite{Song2013} or, more recently, to a combined effect of spin-orbit and pseudo-spin physics\cite{Tuan2014}.
  
Adatoms, in particular,  provide a possible course to engineer spin based effects.
The nature of the SOC, however, might be different for different adatoms. Very light atoms, like H, which are adsorbed on top of a single C atom and have a small intrinsic atomic SOC, introduce SOC to the graphene carriers by distorting the otherwise flat graphene sheet\cite{CastroNeto2009a}. This is due to the  $sp^3$ like structure adopted by the hybridized C atom that induces a local coupling between the carbons $p_z$ and $\sigma$ orbitals. In that case, the effect of the atomic SOC of the C atoms changes (locally) from being a second order effect to  a first order one with the corresponding increase of the effective spin-flip processes in the graphene carriers. A characteristic of this mechanism is that the resulting SOC is proportional to the local lattice distortion.\cite{CastroNeto2009a} On the other hand, more heavy atoms can introduce spin effects by processes that involve their own intrinsic SOC, i.e. where the spin-flip occurs in the adatoms orbitals and not on the C atoms. In the general case both mechanisms are present. 

The fluorine atom on graphene is adsorbed in a top position. It bounds covalently to the C atom below it (we will refer to it as the C$_0$ atom), therefore introducing a local distortion in the graphene lattice. However, owing to its strong electro-negativity, there is a charge transfer from the graphene sheet to the F adatom\cite{Sofo2011, Chan2011,Guzman2014}. This charge transfer to the F adatom can be controlled by changing the carriers concentration of graphene. Associated to this, there is a $sp^3$-$sp^2$ like crossover of the hybridization of the C$_0$ atom \cite{Sofo2011, Guzman2014}. Since this crossover causes a strong reduction of the lattice distortion, one would expect, as suggested by the scenario described in Ref.~[\onlinecite{CastroNeto2009a}],  that the induced SOC associated with it  will also decrease, thus providing a controllable way to reduce the spin relaxation.

Here we show, using Density Functional Theory (DFT) calculations, that, contrary to this expectation, the SOC in dilute fluorinated graphene is strongly enhanced by electron doping. This is due to the fact that the main contribution to the SOC comes from the intrinsic SOC of the F adatom\cite{Irmer2014} and not from the C atoms as the lattice distortion mechanism requires. This result is validated by a simple tight-binding model, constructed using the Wannier functions parameter obtained from the ab initio calculations,  which  accounts  both qualitatively and quantitatively for this effect.

The rest of the work is organized as follows:  DFT results for the band structure and the projected density of states (PDOS) for different geometries are presented in Sec. \ref{DFT}. A tight-binding model based on the calculations of the maximally localized Wannier is introduced in Sec. \ref{TB} to discuss the microscopic mechanism leading to SOC. We finally conclude in Sec. \ref{conclu}.

\begin{figure*}[tb]
\includegraphics[width=.95\textwidth]{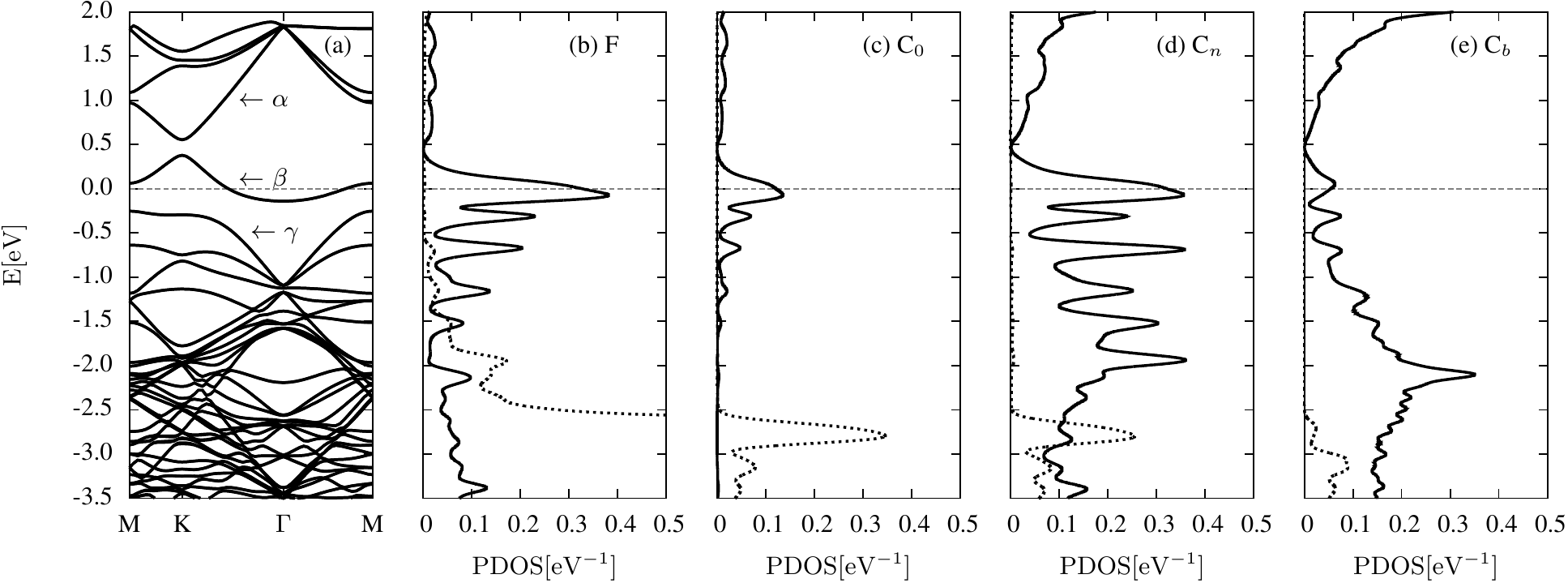}
\caption{ (a) Electronic band structure for a $7\times 7$ supercell containing $98$ C atoms and a single F adatom. The horizontal dashed line indicates the Fermi Energy $\Ef$ which is set to zero. (b-e) projected density of states (PDOS) onto the $p_z$ (solid line) and p$_{x(y)}$ orbitals (dashed line) for the F adatom, the carbon bound to it (C$_0$), one of the three nearest C atoms (C$_n$) of the latter, and a `bulk' (far from the adatom ) carbon atom (C$_b$). $\alpha$, $\beta$ and $\gamma$ label the closest bands to $\Ef$. }
\label{bands_PDOS_7x7}
\end{figure*}
\section{DFT results\label{DFT}}

The DFT calculations were performed with the Quantum Espresso package \cite{QE} employing density functional theory and the Perdew-Burke-Ernzerhoff (PBE) exchange-correlation functional.\cite{pbe} An ultra soft description of the ion-electron interaction was used \cite{Vanderbilt1990} together with a plane-wave basis set for the electronic wave functions and the charge density, with energy cutoffs of $70$ and $420$ Ry, respectively. The electronic Brillouin zone integration was sampled with an uniform k-point mesh ($3\!\times\! 3\!\times\! 1$) and a Gaussian smearing of $0.005$ Ry. The two-dimensional behavior of graphene was simulated by adding a vacuum region of $12$ \AA~ above it. All the structures were relaxed using a criteria of forces and stresses on atoms of $0.005$eV/\AA~ and $0.3$GPa, respectively. The convergence tolerance of energy was set to $10^{-5}$ Ha ($1$ Ha = $27.21$ eV). To correct for the dipole moment generated in the cell and to improve convergence with respect to the periodic cell size, monopole and dipole corrections were considered.\cite{Neugebauer1992} This is particularly important in the doped cases. Doping  of the unit cell (added/removed  electrons) where compensated by an uniformly distributed background charge.

\subsection{Electronic band structure}
Let us first analyze the band structure of fluorinated graphene in the absence of additional doping (neutral case). Since we are interested in describing the SOC introduced by a single impurity (corresponding to a diluted fluorinated graphene sample), for our DFT calculations we use super cells as large as posible (limited by the computational resources). The use of large super cells is also important in the case of F to avoid long range Coulomb interactions among impurities as F adatoms acquire some charge when absorbed on graphene\cite{Sofo2011,Guzman2014,Irmer2014}.   

Figure \ref{bands_PDOS_7x7}(a) shows the DFT electronic band structure for a $7 \times 7$ supercell containing $98$ C atoms and a single F atom. The Fermi energy $\Ef$ has been set to zero, $\Ef=0$. The path in the reciprocal space is labeled using the standard notation for the hexagonal Brillouin zone of the supercell lattice. Though negligible on this scale, each band is split in two due to the SOC---this is analyzed  in detail in the next section.
Since one would naively expect to observe graphene's  Dirac cone at the $K$ point, it is instructive to mention why that is not the case: our DFT supercell contains a single F atom and consequently all impurities in our periodic system are on the same graphene sublattice. This breaks off the sublattice symmetry and opens a gap in the graphene spectrum.\cite{Pereira2008,Cheianov2010} Such gap is proportional to the impurity concentration and hence to the supercell size.  
While this effect should not have any relevance for the calculation of local parameters for a large enough supercell,  it is important to keep it in mind for a proper interpretation of the results---the breaking of the sublattice symmetry can be eliminated from the calculations, at the price of increasing the F concentration or increasing the supercell size, by including two F atoms on the supercell, one on each sublattice.

To analyze the character of the bands, it is useful to look at the weight of the corresponding states on  the atomic orbitals of each type of atom. That is, to look at the projected density of states (PDOS). This is done  in Fig. \ref{bands_PDOS_7x7} for four different atoms in the supercell: F (b), C$_0$ (c), one of the three C$_n$ carbon atoms around $C_0$ (d), and a `bulk' (far from the adatom) C atom (e). The solid line corresponds to the  
 $p_z$ orbitals while the dashed one corresponds to the $p_x(y)$ orbitals---in this energy range the weight on the s orbitals is smaller than $0.02$ eV$^{-1}$. It is apparent from the figure that the central band  (labelled by  $\beta$) and the band below it ($\gamma$) are the ones that contain the impurity states as they  carry a large weight on the $p_z$ orbitals of the F adatom, the C$_0$ atom, and its three nearest neighbors C$_n$ atoms.\cite{Sofo2012}

\begin{figure*}[tb]
\includegraphics[width=.95\textwidth]{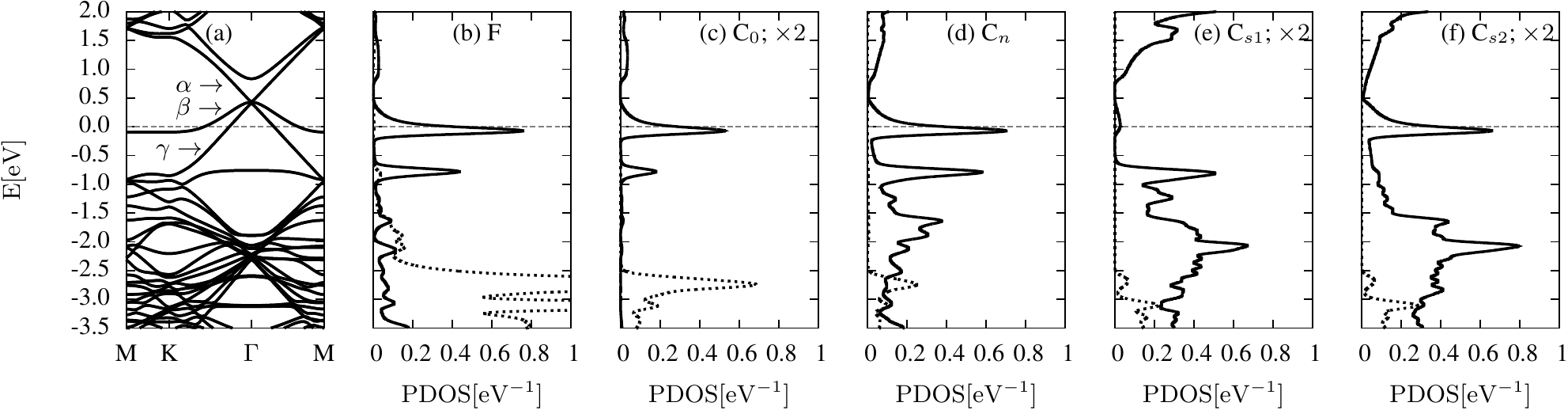}
\caption{Same as Fig. \ref{bands_PDOS_7x7} but for a $6\times 6$ supercell containing $72$ C atoms an a single F adatom. The last rightmost panels correspond to a C atom far from the adatom in the same (e) and the opposite (f) sublattice. Notice that in this case there is a clear Dirac-like band and that the F and the C$_0$ atoms are essentially decoupled from it.}
\label{bands_PDOS_6x6}
\end{figure*}

With the aim of understanding the adatom induced SOC, a very interesting situation occurs for supercells with a single F and sizes of the form $3n\times3n$ with $n$ an integer number. In this case, the perturbation potential induced by the adatom $U(\bm{r})$, assumed to be local, has the right symmetry as to mix the two non-equivalent graphene's Dirac cones. Namely, the Fourier transform of the periodic adatom potential contains non-zero components $U_{\bm{k}}$ with $\bm{k}$ connecting  the two cones. Such admixture leads to a partial decoupling of the graphene $\pi$ band from the adatoms. This  closes the gap induced by  the sublattice symmetry break and restores a single Dirac cone. In addition, the impurity band is more clearly developed. The corresponding band structure is presented in Fig. \ref{bands_PDOS_6x6} for a $6\times6$ supercell, together with the PDOS. Notice that a Dirac cone can be  clearly identified and that in those bands there is essentially no weight of the fluorine's $p_z$ orbital which, in this energy range, is mainly concentrated on the rather flat  impurity band (the $\beta$ band in Fig.~\ref{bands_PDOS_6x6}(a)).
This decoupling, that occurs for a particular geometry, and though rather artificial, will help us to separate different contributions to the SO splitting of the bands.

\subsection{Gate dependence of the spin-orbit splitting.}
We now turn our attention to the splitting of the bands induced by the SOC and, in particular, to the effect of doping on such splitting.

The main contribution to the SO splitting in this system arises from the atomic SOC of the  C and the F atoms. 
The contribution from the C atoms is known to be a second order effect in flat graphene own to the reflexion symmetry of the graphene's plane: symmetry prohibit a direct coupling between the $p_z$ and p$_{x(y)}$ orbitals of  adjacent C atoms---recall that atomic SOC mixes $p$ orbitals on the same lattice site.
This symmetry is locally broken in the presence of an adatom that sits on top a C atom due to the lattice distortion it introduces. This was shown\cite{CastroNeto2009a} to enhance the SOC by breaking the above mentioned selection rule. The resulting SOC is proportional to the distortion of the lattice. 
The latter depends, in the case of F adatoms, on the  doping level of the graphene sheet due to the charge transfer from the graphene to the F adatom\cite{Sofo2012,Guzman2014}. Hence, one expects a strong dependence of the C contribution to the SO splitting upon doping.
Namely, since doping with electrons (holes) reduces (enhances) the $sp^3$ character of the local hybridization of the C$_0$ atom, we expect this contribution to the SO splitting to decrease (increase).

The other source of SOC is the adatom itself. Contrary to the case of H, the SOC in the F atom is not negligible and must be accounted for. How this contribution changes with doping, if it does, is however  not obvious {\it a priori}. 

\begin{figure}[t]
\includegraphics[width=.95\columnwidth]{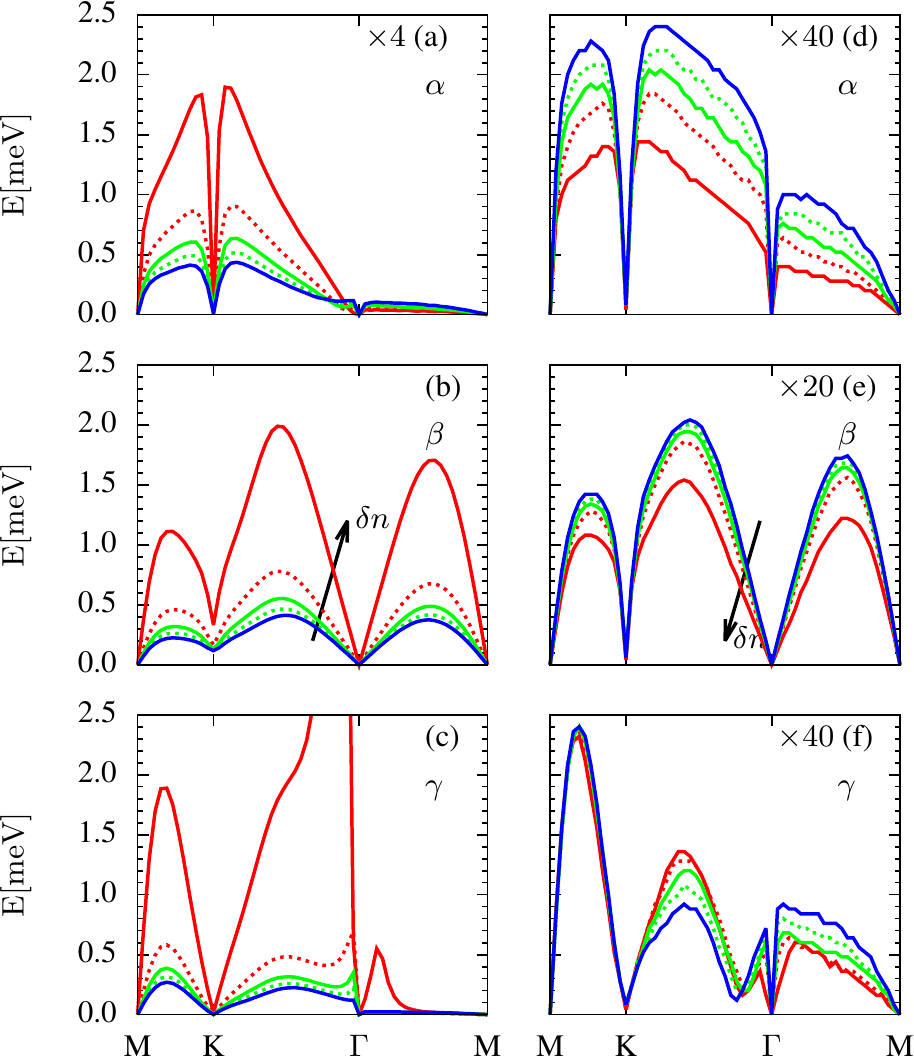}
\caption{(color online) SO splitting of the $\alpha$, $\beta$, and $\gamma$ bands indicated in Fig. \ref{bands_PDOS_7x7}(a) for the cases where the SOC of the F atom is included in the DFT calculation [(a-c)], and those where it is removed [(d-f)] as a function of the doping, $\delta n=0,\pm1/2,\pm1$ electrons per unit cell. The arrows indicate the direction of increment of $\delta n$ while the multiplication factors represent the scaling used in the energy scale. Notice that the doping dependence is just the oposite with and without SOC in the F atom.}
\label{SO_spliting}
\end{figure}

The DFT results for the SO splitting of  three bands for the case of the $7\times7$ supercell (indicated as $\alpha$, $\beta$, and $\gamma$ in Fig.~\ref{bands_PDOS_7x7}(a))  is shown in Figs. \ref{SO_spliting}(a-c) for five different doping concentrations: $\delta n=0,\pm1/2,\pm1$ is the number of additional electrons per unit cell (hence $\delta n=0$ corresponds to the neutral case).
We find that \textit{the SO splitting changes for all the bands, being enhanced by electron doping and reduced by hole doping}.
This is just the opposite behavior that is expected  based on the distortion induced SOC.  We must then conclude that the main contribution to the observed splitting is not the atomic SO of the C atoms but the one coming from the F adatoms.\cite{Irmer2014}
To verify this, we repeated our calculation but without including the SOC on the F---this is done by using a scalar relativistic pseudo-potential for the F atom as implemented in the Quantum Espresso code\cite{QE}). The results are shown in Figs. \ref{SO_spliting}(d-f). There are two important points to notice: (i) the magnitude of the splitting is reduced by a factor of approximately five, consistent with the difference in magnitude of the atomic SO between F and C ($\sim50$ and $\sim10$ meV, respectively), and signaling that the main source of SO is absent; (ii) the SO splitting shows now the expected behavior as a function of doping for a deformation induced SO: it is reduced by electron doping and enhanced by hole doping.

\begin{figure}[t]
\includegraphics[width=.95\columnwidth]{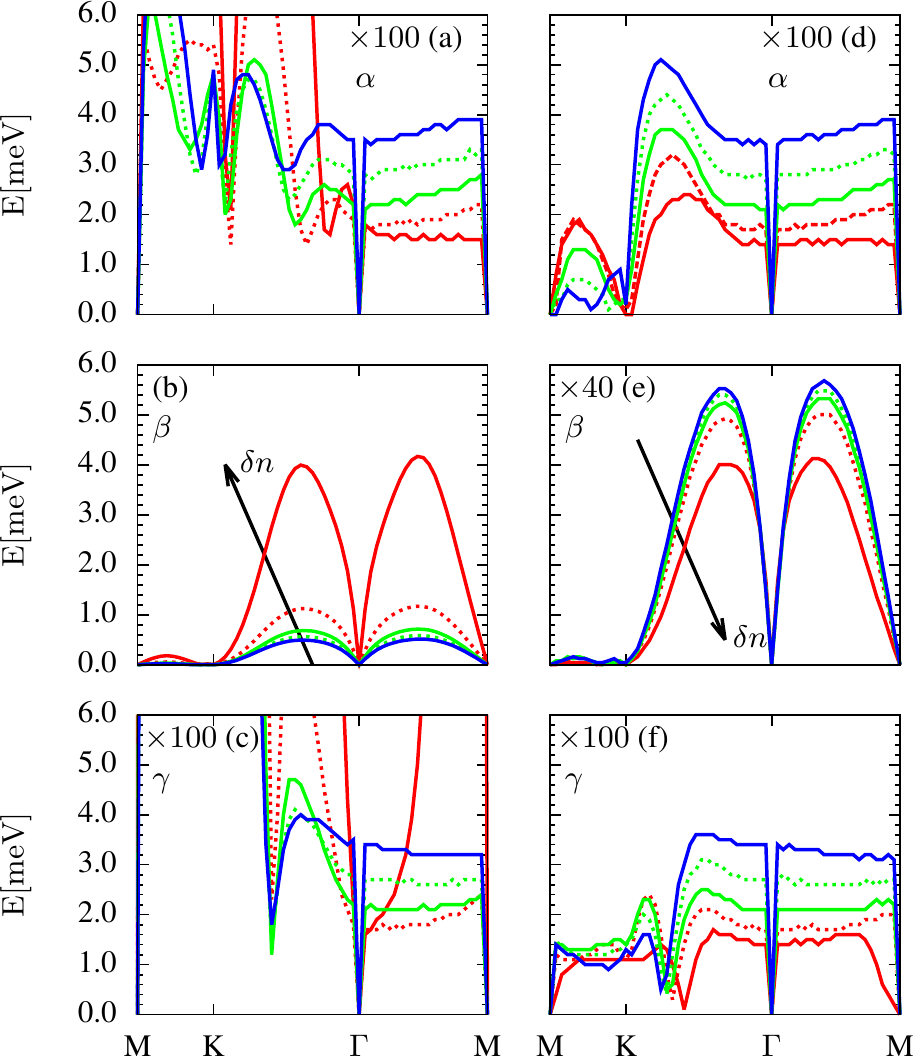}
\caption{(color online) SO splitting as in Fig. \ref{SO_spliting} but for the $6\times6$ supercell. Notice that only the impurity band ($\beta$) maintains the doping dependence of the F induced SO splitting, the rest behave as the C induced SOC near the $\Gamma$ point.}
\label{SO_spliting_6x6}
\end{figure}

To further verify the origin of the SO splitting in fluorinated graphene we have done the calculation keeping the SOC in the F atom but using a $6\times6$ supercell that leads to the decoupling of the $\pi$ band from the adatoms as discussed in the previous section.
Quite interestingly, the bands show now different behaviors (Figs. \ref{SO_spliting_6x6}(a-c)): those with a small  weight on the fluorine's $p_z$ orbitals (bands $\alpha$ and $\gamma$) have a SO splitting compatible in magnitude with a deformation induced SO and the corresponding dependence with doping. On the other hand, the band with a large weight on the F ($\beta$) shows a large SO splitting and the opposite doping dependence. Again, this is confirmed by removing the SOC in F as shown in Figs. \ref{SO_spliting_6x6}(d-f).

\section{Tight-binding model\label{TB}}
\begin{figure}[t]
\includegraphics[width=.85\columnwidth]{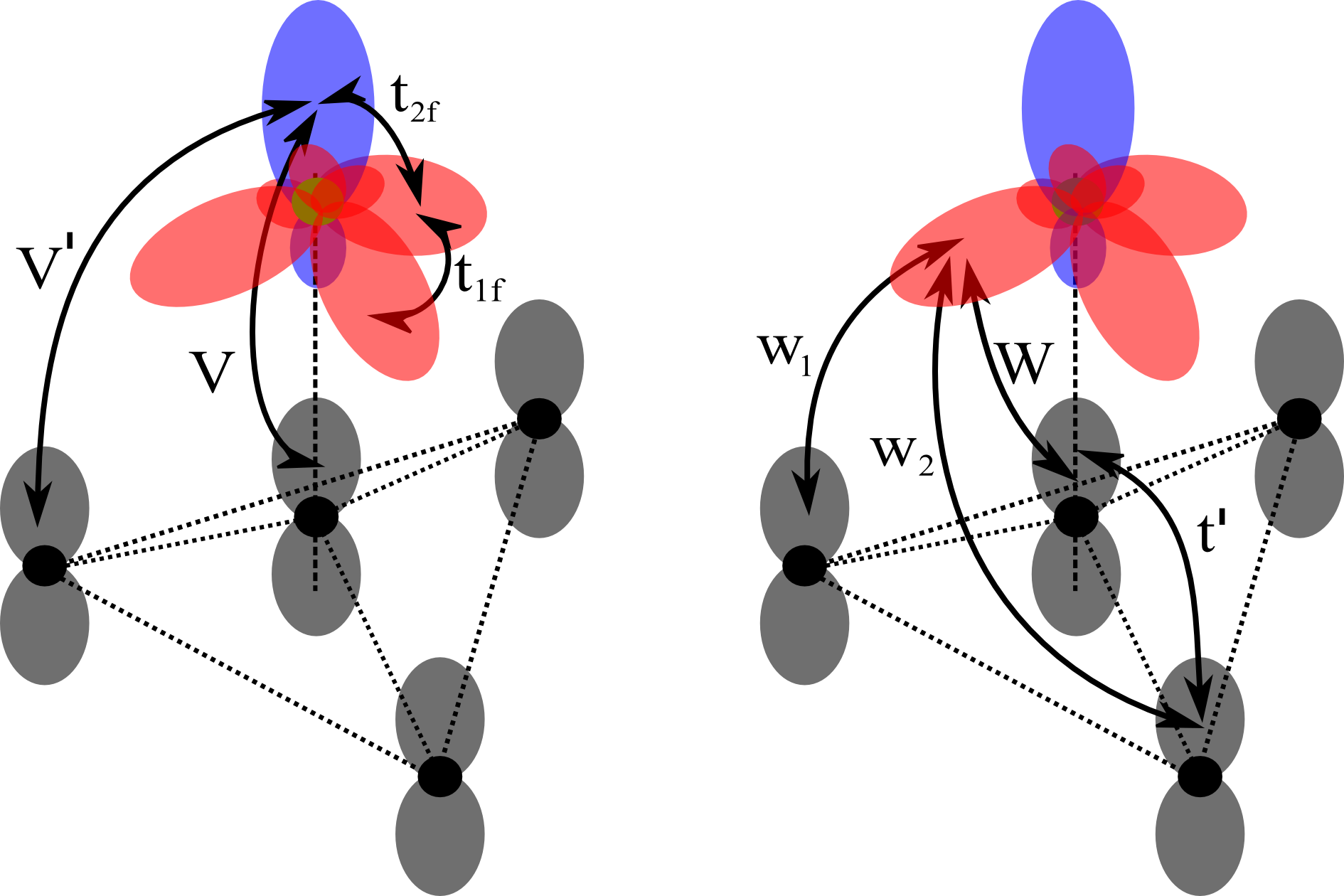}
\caption{(color online) Schematic representation of the hybridized orbitals used in the tight-binding model with their hopping matrix elements.}
\label{scheme}
\end{figure}

The above results clearly demonstrate that in fluorinated graphene the main source of SOC are the F adatoms, in agreement with Ref. [\onlinecite{Irmer2014}].  However, the origin of its strong dependence on doping is not at all clear. To better understand the underlying microscopic mechanism we now construct a single particle tight-binding model that includes the most relevant orbitals, and later on the atomic SOC of F.
Namely, we consider a small cluster consisting of one F and four C atoms (see Fig. \ref{scheme}) embedded in graphene. The Hamiltonian is given by $\Ha=\Ha_{\mathrm{F}}+\Ha_{\mathrm{C}}+\Ha_{\mathrm{FC}}$. The first term describes the isolated F atom
\begin{equation}
\Ha_{\mathrm{F}}=\sum_{\xi,\sigma}\varepsilon_p p^{\dagger}_{\xi\sigma}p_{\xi\sigma}^{}+\sum_{\sigma}\varepsilon_s s^{\dagger}_{\sigma}s_{\sigma}^{}\,,
\label{HF}
\end{equation}
where $p^{\dagger}_{\xi\sigma}$ creates an electron with spin $\sigma$ in the $2$p$_{\xi}$ orbital ($\xi=x, y, z$) and $s^{\dagger}_{\sigma}$ creates an electron with spin $\sigma$  in the $2$s orbital. 
Given the symmetry of the system, it is convenient, for the definition of the hopping parameters and for the sake of comparison with the Wannier functions described below, to work in a hybridized  basis. To this end we define the following creation operators,\cite{CastroNeto2009a} 
\begin{eqnarray}
\nonumber
f^{\dagger}_{z\sigma}&=&A\,s^{\dagger}_{\sigma}+\sqrt{1-A^2}\,p^{\dagger}_{z\sigma}\,,\\
\nonumber
f^{\dagger}_{1\sigma}&=&\frac{\sqrt{1-A^2}}{\sqrt{3}}s^{\dagger}_{\sigma}-\frac{A}{\sqrt{3}}\,p^{\dagger}_{z\sigma}+\sqrt{\frac{2}{3}}p^{\dagger}_{x\sigma}\,,\\
f^{\dagger}_{2\sigma}&=&\frac{\sqrt{1-A^2}}{\sqrt{3}}s^{\dagger}_{\sigma}-\frac{A}{\sqrt{3}}\,p^{\dagger}_{z\sigma}-\frac{1}{\sqrt{6}}p^{\dagger}_{x\sigma}+\frac{1}{\sqrt{2}}p^{\dagger}_{y\sigma}\,,\\
\nonumber
f^{\dagger}_{3\sigma}&=&\frac{\sqrt{1-A^2}}{\sqrt{3}}s^{\dagger}_{\sigma}-\frac{A}{\sqrt{3}}\,p^{\dagger}_{z\sigma}-\frac{1}{\sqrt{6}}p^{\dagger}_{x\sigma}-\frac{1}{\sqrt{2}}p^{\dagger}_{y\sigma}\,.
\end{eqnarray}
The parameter $A$ serves to interpolate between two extreme cases: (i) $A=0$, in which case $f^{\dagger}_{z\sigma}=p^{\dagger}_{z\sigma}$ while the other orbitals, $f^{\dagger}_{i\sigma}$, correspond to the standard $sp^2$ hybrid orbitals; (ii) $A=1/2$, where all orbitals correspond to the $sp^3$ hybridization, with one of them pointing in the $z$ direction.
In terms of these orbitals,  the fluorine Hamiltonian takes the form
\begin{eqnarray}
\nonumber
\Ha_{\mathrm{F}}&=&\sum_{\sigma}\varepsilon_{z}\, f^{\dagger}_{z\sigma}f_{z\sigma}^{}+\sum_{i,\sigma} \varepsilon_f\, f^{\dagger}_{i\sigma}f_{i \sigma}^{}+\sum_{i\ne j,\sigma} t_{1f}\,  f^{\dagger}_{i\sigma}f_{j\sigma}^{}\\
&&+\sum_{i,\sigma} t_{2f}\, \left(f^{\dagger}_{z\sigma}f_{i \sigma}^{}+f^{\dagger}_{i\sigma}f_{z \sigma}^{}\right)\,,
\label{Hf}
\end{eqnarray}
where $i,j=1,2,3$, and
\begin{eqnarray}
\label{param}
\nonumber
\varepsilon_z&=&\left(1-A^2\right) \ve_p+A^2\ve_s\,,\quad \varepsilon_f=\frac{\left(1-A ^2\right)}{3}\varepsilon_{s}+\frac{\left(2+A^2\right)}{3}\varepsilon_{p}\,,\\
t_{1f}&=&\frac{\left(1-A^2\right)}{3}\left(\varepsilon_{s}-\varepsilon_{p}\right)\,,\quad t_{2f}=\frac{A\sqrt{1-A^2}}{\sqrt{3}}\left(\varepsilon_{s}-\varepsilon_{p}\right)\,.
\end{eqnarray}
The Hamiltonian of the C atoms is
\begin{eqnarray}
\nonumber
\Ha_{\mathrm{C}}&=&\sum_\sigma\varepsilon_0\, c^{\dagger}_{0\sigma}c_{0\sigma}^{}+\sum_{i,\sigma}\varepsilon_1\, c^{\dagger}_{i\sigma}c_{i\sigma}^{}+t'\sum_{i,\sigma} \left(c^{\dagger}_{0\sigma}c_{i\sigma}^{}+ c^{\dagger}_{i\sigma}c_{0\sigma}^{}\right)\\
&&+t'_2\sum_{i\ne j,\sigma} c^{\dagger}_{i\sigma}c_{j\sigma}^{}+\Ha_g\,,
\end{eqnarray}
here $c^{\dagger}_{0\sigma}$ and $c^{\dagger}_{i\sigma}$ ($i=1,2,3$) create electrons with spin $\sigma$ at the $p_{z}$ orbitals of the central C atom (C$_0$) and  the side carbon atoms (C$_n$), respectively, and $\Ha_g$ is the Hamiltonian of the rest of the $p_z$ orbitals of the graphene sheet with energy $\varepsilon_g$ and hopping $t$.
Finally, $\Ha_{\mathrm{FC}}$  includes the  hybridization between the F and C orbitals,
\begin{eqnarray}
\Ha_{\mathrm{FC}}&=&V\sum_{\sigma} \left(f^{\dagger}_{z\sigma}c_{0\sigma}^{}+ c^{\dagger}_{0\sigma}f_{z\sigma}^{}\right)+V'\sum_{i\sigma} \left(f^{\dagger}_{z\sigma}c_{i\sigma}^{}+ c^{\dagger}_{i\sigma}f_{z\sigma}^{}\right)\\
\nonumber
&&+W\sum_{i,\sigma} \left(f^{\dagger}_{i\sigma}c_{0\sigma}^{}+ c^{\dagger}_{0\sigma}f_{i\sigma}^{}\right)
+\sum_{ij,\sigma} w_{ij} \left(f^{\dagger}_{i\sigma}c_{j\sigma}^{}+ c^{\dagger}_{j\sigma}f_{i\sigma}^{}\right)\,,
\end{eqnarray}
where  $w_{ij}$ takes only two values: $w_1$ if the $f_i^\dagger$ orbital `points' towards the $c_j^\dagger$ orbital and $w_2$ otherwise (see Fig. \ref{scheme}). 

The next step is to properly estimate all the parameters: energies of the hybridized orbitals and hopping matrix elements. 
This can be achieved with the help of the Wannier90 code\cite{Mostofi2008} as shown in the next section.
\subsection{Wannier functions\label{WannierSec}}
A simple way to build a tight-binding model from the DFT results is  to find the maximally localized Wannier functions\cite{Marzari2012} that describe  the DFT band \textit{exactly} on the energy range of interest (around $\Ef$, for instance) and  the rest of the spectrum only approximately, depending on the number of orbitals used in the calculations.
To this end, we use the Wannier90 code in a $4\times4$ supercell. While our calculations include the $\sigma$ orbitals between the C atoms, required to properly describe the DFT band structure,  here we present only those parameters that are relevant for the simplified tigh-binding Hamiltonian described above which is enough to capture the SO splitting of the energy bands around $\Ef$. 
Table \ref{tabla1} presents such parameters  for several doping configurations. 
\begin{table}[tb]
\begin{ruledtabular}
\begin{tabular}{lrrrrr}
Doping & $-1$&$-\frac{1}{2}$& neutral & $+\frac{1}{2}$&$+1$\\
\hline
$\ve_z$&-10.43&-9.63 &-8.72 &-7.50&-5.40 \\
$\ve_f$&-12.29&-11.45&-10.35 &-8.86&-6.27 \\
$\ve_0$&-0.08&0.25&0.38 &0.31&0.36 \\
$\ve_1$&-2.55&-1.91 &-1.30 &-0.80&-0.27 \\
$t'$ &-2.45&-2.37 &-2.31 &-2.34&-2.71\\
$t_2'$ &-0.07&-0.09 &-0.11 &-0.09&0.13\\
$t_{1f}$&-5.34 &-5.25&-5.15 &-5.02&-4.85 \\
$t_{2f}$&-3.99&-4.07 &-4.17 &-4.25&-4.33 \\
$V$&3.34& 3.31&3.11 &2.48&1.11\\
$V'$ &-0.21&-0.18 &-0.14 &-0.06&0.09\\
$W$&-3.47&-3.43 &-3.18 &-2.42&-0.88\\
$w_1$&-0.54 &-0.56&-0.56&-0.53&-0.42 \\
$w_2$&0.30&0.28&0.25&0.20 &0.07 \\
$\ve_g$ &-2.08&-1.53&-0.99&-0.52 &-0.34 \\
$t$&-2.85&-2.86&-2.87&-2.88 &-2.90 
 \end{tabular}
 \end{ruledtabular}
\caption{Tight-binding parameters (in eV) calculated by the Wannier90 program for different doping concentrations (number of additional electrons per unit cell) in a $4\times4$ supercell. For a description of the parameters see Fig.~\ref{scheme}.}
\label{tabla1}
\end{table}
\begin{table}[b]
\begin{ruledtabular}
\begin{tabular}{cccccc}
Doping & $\ve_s$ [eV] & $\ve_p$ [eV] & $A$&$t_{2f}^\mathrm{calc}$ [eV] & $t_{2f}^\mathrm{calc}/t_{2f}$\\
\hline
$-1$ & $-26.45$ & $-6.95$ & $0.42$ & $-4.31$ & $1.08$\\
$-\frac{1}{2}$ & $-25.38$ & $-6.20$ & $0.42$& $-4.24$ & $1.04$\\
neutral & $-24.17$ & $-5.20$ & $0.43$ & $-4.26$ & $1.02$\\
$+\frac{1}{2}$ & $-22.56$ & $-3.84$ & $0.44$& $-4.29$ & $1.01$\\
$+1$ & $-19.95$ & $-1.42$ & $0.46$ & $-4.39$& $1.01$
 \end{tabular}
 \end{ruledtabular}
\caption{Parameters of the model Hamiltonian $\Ha_\mathcal{F}$. $\ve_s$, $\ve_p$, and $A$ are calculated by fitting the values of $\ve_z$, $\ve_f$, and $t_{1f}$  obtained by the Wannier functions method. The value of $t_{2f}^\mathrm{calc}$ is derived from the former, being consistent with the Wannier calculation.}
\label{tabla2}
\end{table}

\begin{figure}[t]
\includegraphics[width=.75\columnwidth]{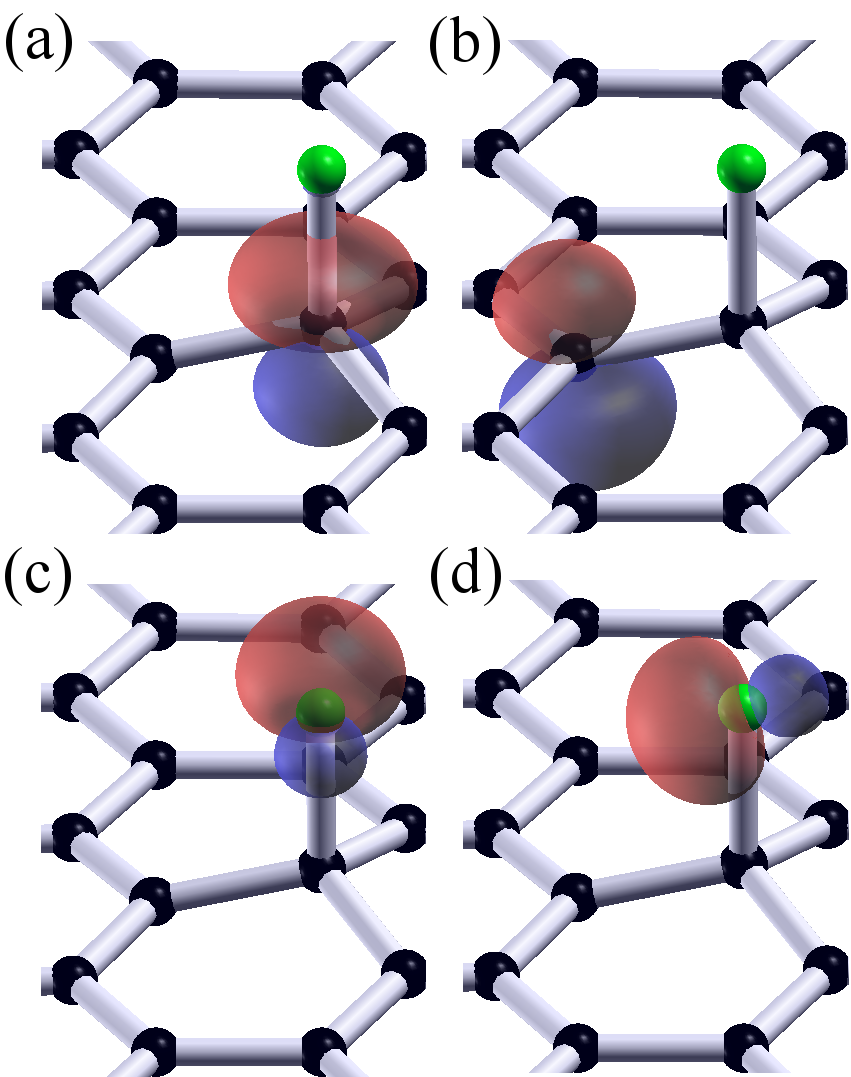}
\caption{(color online) Maximally localized Wannier functions as obtained with the Wannier90 code and used for the tight-binding model. They  correspond to: (a)  $c_0^\dagger$, (b) one of the $c_i^\dagger$ orbitals, (c) $f_z^\dagger$, and (d) one of the $f_i^\dagger$ orbitals.}
\label{Wannier}
\end{figure}

The maximally localized Wannier functions obtained for the  $c_0^\dagger$, $c_i^\dagger$, $f_z^\dagger$, and $f_i^\dagger$ orbitals are shown in Fig. \ref{Wannier}. It is apparent from the figure that the F orbitals can be interpreted as the hybridized atomic orbitals described above while the C$_0$ and C$_n$ orbitals need to be considered as `effective' orbitals (the former being more $sp^3$-like and the latter as a tilted $p_z$-like one). 

The interpretation of the F orbitals as hybridized atomic orbitals is also supported by the following: our tight-binding model has three parameters ($\ve_s$, $\ve_p$, $A$) from where four parameters are obtained [cf. Eqs. (\ref{param})]. These four parameters are calculated independently in the Wannier90. Hence, we can use three of them, say ($\ve_z$, $\ve_f$, $t_{1f}$), to calculate ($\ve_s$, $\ve_p$, $A$) and from them the value of $t_{2f}^\mathrm{calc}$ to be compared with the value $t_{2f}$ shown in Table \ref{tabla1}. Such comparison, together with the values obtained for $\ve_s$, $\ve_p$, and $A$ is shown in Table \ref{tabla2}.
Our results show that, within this model, the main effect of doping on the F's orbitals is to shift their energy  up as more electrons are added to the supercell and to slightly change  the hybridization parameter $A$  ($<10\%$).
This energy shifting is to be expected from a naive mean field argument as doping induces a charging of the F adatom.\cite{Sofo2011,Guzman2014} We would like to point out that while with this very simple model one cannot expect to fit the value of $t_{2f}$, the calculated hopping term  ($t_{2f}^\mathrm{calc}$) is quite reasonable and follow the overall trend with doping---except for the extreme hole doping case.

It is worth noticing that  while there is an overall shift of all energies with doping, the important magnitude is their relative shift, which is by far  larger in the F atom. This is the reason for the change of the SOC as we discuss below.

Finally, we emphasize that the present calculation was done in a $4\times4$ supercell which is relatively small as to avoid the finite size effects related to the long range Coulomb repulsion between F adatoms\cite{Guzman2014,Irmer2014}---in our case the size is limited by  the heavy computational resources needed for the calculation of the Wannier orbitals. Therefore, we may expect that our tight-binding parameters will not be reliable for large electron doping, where the finite size effect are stronger. Yet, we will show that they provide a good description of the SO splitting of the bands.

\begin{figure}[t]
\includegraphics[width=.9\columnwidth]{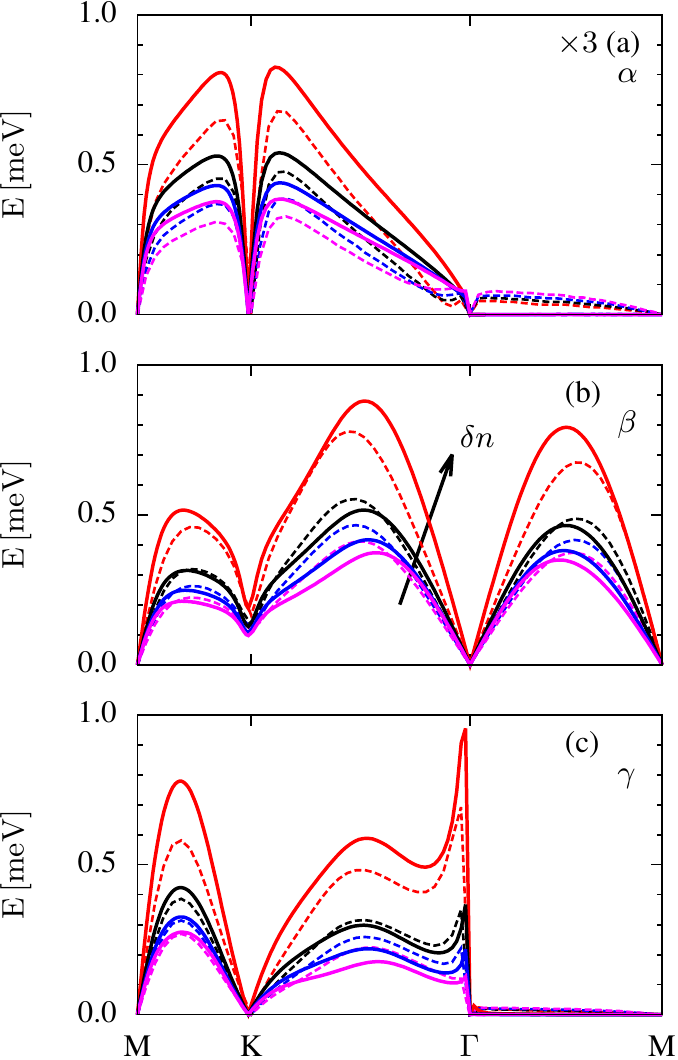}
\caption{(color online) Comparison of the SO splitting of the $\alpha$, $\beta$, and $\gamma$ bands obtained with the simplified tight-binding model (solid lines) and the DFT results of Fig. \ref{SO_spliting} (dashed lines), excluding the case of $+1$ electron doping. To obtain a good agreement we use $\alpha=40$meV while the rest of the tight-binding parameters are taken from Table \ref{tabla1}. }
\label{compa}
\end{figure}
\subsection{Spin-orbit coupling}

As our DFT results show that the main source of SOC comes from the F adatom, we will include in our tight-binding model only the atomic SOC of the F---an extension to include the effect of the SOC on the C atoms is straightforward. Such term is of the form $
\Ha_{\mathrm{F}}^{\mathrm{SO}}=\alpha\, \bm{L} \cdot\bm{S}$, 
with $ \bm{L}$ and $ \bm{S}$  the orbital angular momentum and spin operators of electrons in the p orbitals, respectively, and  the SO parameter $\alpha\sim50$meV for F.\cite{Radford1961,Irmer2014} In the $\{$p$_{x\up}$, p$_{y\up}$, p$_{z\up}$, p$_{x\dn}$, p$_{y\dn}$, p$_{z\dn}$$\}$ basis it takes the following matrix form\cite{Petersen2000,Ast2012}
\begin{equation}
\Ha_{\mathrm{F}}^{\mathrm{SO}}=\frac{\alpha}{2}\left(
\begin{tabular}{cccccc}
$0$&$-i$&$0$&$0$&$0$&$1$\\
$i$&$0$&$0$&$0$&$0$&$-i$\\
$0$&$0$&$0$&$-1$&$i$&$0$\\
$0$&$0$&$-1$&$0$&$i$&$0$\\
$0$&$0$&$-i$&$-i$&$0$&$0$\\
$1$&$i$&$0$&$0$&$0$&$0$\\
\end{tabular}\right)\,.
\end{equation}
By adding $\Ha_{\mathrm{F}}^{\mathrm{SO}}$ to the tight-binding Hamiltonian presented in the previous section we can now compute the splitting of the $\alpha$, $\beta$ and $\gamma$ bands for a given supercell and compare it with the DFT data. This is done in Fig. \ref{compa} using the parameters of Table \ref{tabla1} and taking $\alpha\sim 40-50$meV.
There are several points to emphasize: (i) there is an overall good qualitative agreement, which further corroborates our model; (ii) the value of the splitting and its behavior with doping is the correct one, indicating that the F adatom is the dominant source of SOC; (iii) our model overestimate the effect of electron doping (the $+1$ case it is strongly overestimated hence is not shown in the figure). Presumably this is due to the finite size  effect discussed in the previous section that affects our tight-binding parameters; (iv) the central band ($\beta$) is the one that is better described, as could be expected.

\begin{figure}[tb]
\includegraphics[width=.7\columnwidth]{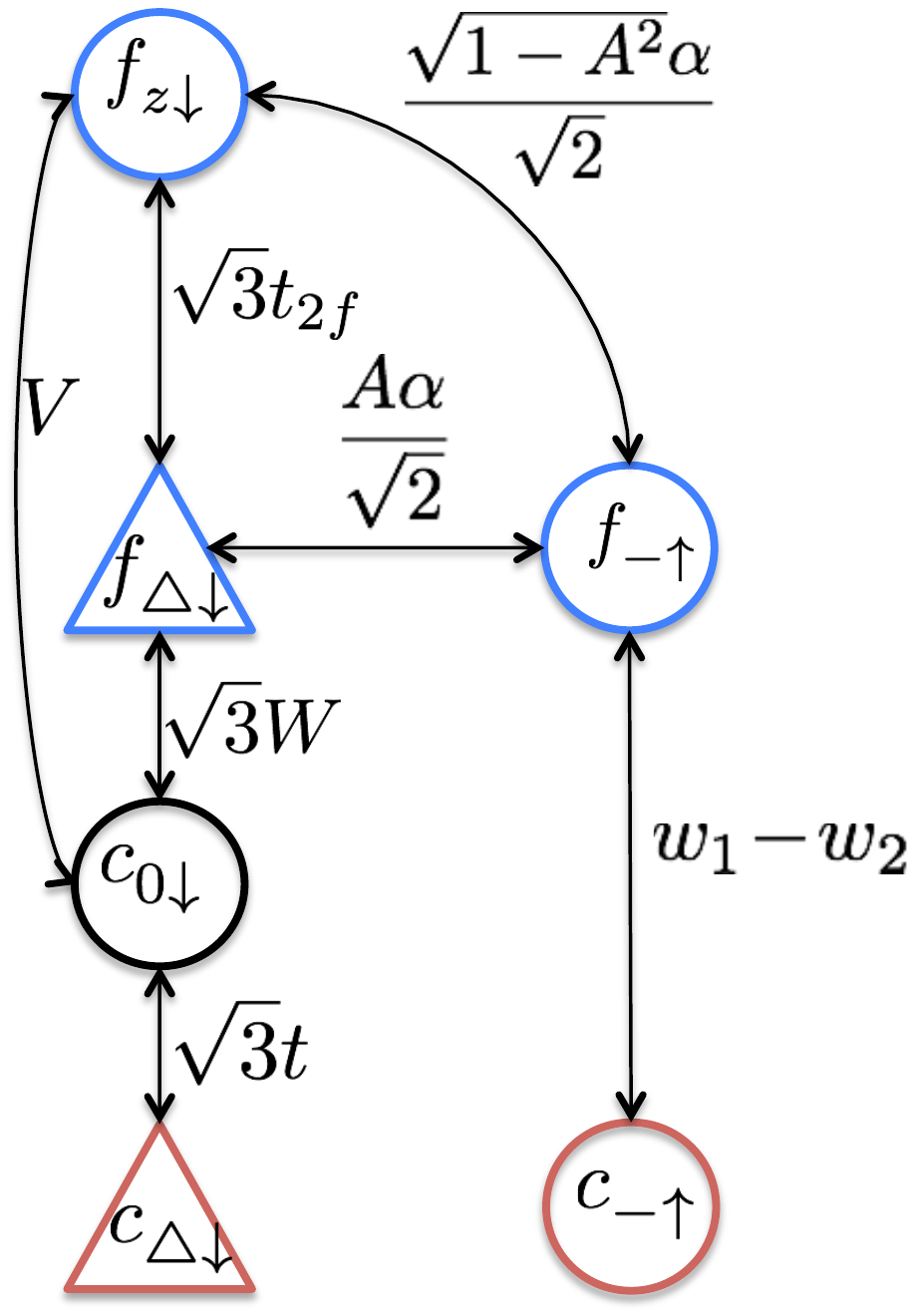}
\caption{(color online) Scheme showing the most relevant hopping parameters of Hamiltonian $\Ha$. The orbital states decoupled by symmetry only couple through the spin flip-term of the fluorine's SOC.}
\label{scheme-H}
\end{figure}

To better understand what are the microscopic spin-flip  processes that lead to the splitting of the bands it is useful to take full advantage of the symmetry of the system  around the adatom and define new symmetrized orbitals as follows, 
\begin{equation}
f^{\dagger}_{\triangle\sigma}=\frac{1}{\sqrt{3}}\sum_{i=1}^3 f^{\dagger}_{i\sigma}\,,\qquad c^{\dagger}_{\triangle\sigma}=\frac{1}{\sqrt{3}}\sum_{i=1}^3 c^{\dagger}_{i\sigma}\,,
\end{equation}
and
\begin{eqnarray}
\nonumber
f^{\dagger}_{+\sigma}&=&\frac{\beta f^{\dagger}_{1\sigma}-f^{\dagger}_{2\sigma}+\beta^*f^{\dagger}_{3\sigma}}{\sqrt{3}}\,,\qquad
f^{\dagger}_{-\sigma}=\frac{\beta^2f^{\dagger}_{1\sigma}+f^{\dagger}_{2\sigma}+\beta^{*2}f^{\dagger}_{3\sigma}}{\sqrt{3}}\,,\\
\nonumber
c^{\dagger}_{+\sigma}&=&\frac{\beta c^{\dagger}_{1\sigma}-c^{\dagger}_{2\sigma}+\beta^*c^{\dagger}_{3\sigma}}{\sqrt{3}}\,,\qquad
c^{\dagger}_{-\sigma}=\frac{\beta^2c^{\dagger}_{1\sigma}+c^{\dagger}_{2\sigma}+\beta^{*2}c^{\dagger}_{3\sigma}}{\sqrt{3}}\,,\\
\end{eqnarray}
where $\beta=\exp(i\pi/3)$. Note that $f^{\dagger}_{\triangle\sigma}=\sqrt{1-A^2}\,s^{\dagger}_{\sigma}-A\,p^{\dagger}_{z\sigma}$ is the combination of the $s$ and $p_z$ orbitals that  is orthogonal to $f^\dagger_{z\sigma}$. In this basis, the full Hamiltonian decouples into orthogonal subspaces with  different spatial symmetry. The key point is that this orthogonal spaces only couple through the splin-flip term of the fluorine's SOC.  The Hamiltonian of the rest of the graphene sheet can also be separated in two sectors with different spatial symmetry---each one of these couple to one of the $c^{\dagger}_{\triangle}$ or $c^{\dagger}_{\pm}$ orbitals.

A scheme of the hierarchy of the hoppings matrix elements in $\Ha$ is shown in Fig. \ref{scheme-H}  for one of the two spin sectors---the other one is analogous but with the time reversal partners. 
Since first order (direct) spin-flip processes only occur between the $f^{\dagger}_{z\sigma}$, $f^{\dagger}_{\triangle\sigma}$  and the $f^{\dagger}_{\pm\bar{\sigma}}$  orbitals of the F adatom, any effective  SOC between  C atoms must be induced by virtual processes that involve them at intermediate steps. This is the way a spin-splitting is induced in all bands. The scheme  makes evident  a couple of things: (i) no virtual spin-flip process is allowed between the $F$ and the C$_0$ atoms as they belong to the same spin subspace, in agreement with the analysis of Ref. [\onlinecite{Irmer2014}]; (ii) virtual spin-flip processes between C atoms, necessarily involve the mixing of the subspaces with opposite spatial symmetry and hence the coupling between the $f^{\dagger}_{\pm\sigma}$ and the $c^{\dagger}_{\pm\sigma}$ orbitals is crucial. The effect of doping enters by changing (reducing or increasing) the energy difference between the F and the C orbitals, making  the virtual spin-flip processes more or less effective.

Finally, it is useful to estimate from our model the effective spin-flip coupling between the C$_n$ atoms around C$_0$ (\textit{ie}, an effective term of the form $\Lambda\, c^\dagger_{1\up}c^{}_{2\dn}+\Lambda^*\, c^\dagger_{2\dn}c^{}_{1\up}$, for instance) in order to compare with other models in the literature\cite{CastroNeto2009a,Irmer2014}. Following the standard procedure  to eliminate the intermediate states using Green functions, we get $\Lambda\sim 2$meV for the neutral case. This effective SOC changes with doping:   it decreases up to roughly $20\%$ for hole doping and increases up to $60\%$ for the cases shown in Fig. \ref{compa}.
This parameter is related to the parameter $\Lambda^\mathrm{B}_\mathrm{PIA}$ introduced in Ref. [\onlinecite{Irmer2014}]. Our result has the same order of magnitude, indicating a large SOC induced by the adatom, but differs in a factor $\sim3$.

\section{Summary \label{conclu}}
We have analyzed the mechanisms governing the appearance of SOC in fluorinated graghene. Our results allow to disentangle the different contributions of the SOC induced on the graphene carriers and to make predictions on its evolution with electron and hole doping.
DFT calculations of a F atom adsorbed on different graphene supercells show significant spin splittings of the low energy bands and give a first indication of the dominance of the atomic SO of the adatom in this case.
The DFT results, combined with the Wannier90 code, allow finding the maximally localized Wannier functions that correctly describe the bands on the energy range of interest to build an effective tight-binding model that includes  effective $\pi$-orbitals of the C atoms and the $2s$ and $2p$ orbitals of the F atom. The so obtained tight-binding model fits the spin-splitting of the low energy bands and allow for a microscopic interpretation of the origin of the SO effects and its doping dependence.
We conclude that:
The SOC in fluorinated graphene is dominated by the intrinsic SO interaction in the F atoms, in agreement with the results of Ref [\onlinecite{Irmer2014}].
The charge transfer from the graphene to the F orbitals has an important effect on the final spin-flip coupling of the graphene carriers. This is mainly due to the charge transfer induced energy shifts of the fluorine $sp$ orbitals that are mixed by the SOC (orbitals with energies $\epsilon_{z}$ and $\epsilon_{f}$ in Hamiltonian (\ref{Hf})).
As this charge transfer can be controlled by doping, the final SO effects can be controlled by gating the sample. While hole doping produces a small decrease of the SOC, electron doping can produce a significant increase of the effect.

It is worth noting that our results show that the spin-splitting at the $K$ point of the Brillouin zone is always extremely small. This is due to the symmetries of the supercells used in the calculation and by no means implies that SO effects on the low energy states of samples with a random distribution of F impurities will be negligible. To see this, consider the single impurity case of the system described by Hamiltonian of Eq. (\ref{Hsym}). There, the important SO parameter is the atomic SOC of the F atom $\alpha$. As discussed in the previous section, this parameter leads to an effective spin-flip coupling between the C atoms, an effect that can be described by an effective Hamiltonian including spin-flip processes in the graghene carriers around the impurity site.\cite{Irmer2014} The spin-flip cross section for this effective impurity model was calculated in Ref. [\onlinecite{CastroNeto2009a}]. An important result of our work is to show that, for fluorine impurities, the spin-flip coupling between the C$_n$ atoms ( $\Lambda$) is much larger than what we could expect from the atomic SOC of the C atoms and, even more important, that its magnitude  can be controlled by gating the sample.

\section{Acknowledgements}
We thank J. Sofo for useful discussions and acknowledge financial support from PICT
Bicentenario 2010-1060 and PICT 2012-379 from ANPCyT, PIP 11220110100832 from CONICET. ADHN and GU acknowledges support from the ICTP associateship program. GU also thanks the Simons Foundation.


\begin{thebibliography}{54}%
\makeatletter
\providecommand \@ifxundefined [1]{%
 \@ifx{#1\undefined}
}%
\providecommand \@ifnum [1]{%
 \ifnum #1\expandafter \@firstoftwo
 \else \expandafter \@secondoftwo
 \fi
}%
\providecommand \@ifx [1]{%
 \ifx #1\expandafter \@firstoftwo
 \else \expandafter \@secondoftwo
 \fi
}%
\providecommand \natexlab [1]{#1}%
\providecommand \enquote  [1]{``#1''}%
\providecommand \bibnamefont  [1]{#1}%
\providecommand \bibfnamefont [1]{#1}%
\providecommand \citenamefont [1]{#1}%
\providecommand \href@noop [0]{\@secondoftwo}%
\providecommand \href [0]{\begingroup \@sanitize@url \@href}%
\providecommand \@href[1]{\@@startlink{#1}\@@href}%
\providecommand \@@href[1]{\endgroup#1\@@endlink}%
\providecommand \@sanitize@url [0]{\catcode `\\12\catcode `\$12\catcode
  `\&12\catcode `\#12\catcode `\^12\catcode `\_12\catcode `\%12\relax}%
\providecommand \@@startlink[1]{}%
\providecommand \@@endlink[0]{}%
\providecommand \url  [0]{\begingroup\@sanitize@url \@url }%
\providecommand \@url [1]{\endgroup\@href {#1}{\urlprefix }}%
\providecommand \urlprefix  [0]{URL }%
\providecommand \Eprint [0]{\href }%
\providecommand \doibase [0]{http://dx.doi.org/}%
\providecommand \selectlanguage [0]{\@gobble}%
\providecommand \bibinfo  [0]{\@secondoftwo}%
\providecommand \bibfield  [0]{\@secondoftwo}%
\providecommand \translation [1]{[#1]}%
\providecommand \BibitemOpen [0]{}%
\providecommand \bibitemStop [0]{}%
\providecommand \bibitemNoStop [0]{.\EOS\space}%
\providecommand \EOS [0]{\spacefactor3000\relax}%
\providecommand \BibitemShut  [1]{\csname bibitem#1\endcsname}%
\let\auto@bib@innerbib\@empty
\bibitem [{\citenamefont {Castro~Neto}\ \emph
  {et~al.}(2009{\natexlab{a}})\citenamefont {Castro~Neto}, \citenamefont
  {Guinea}, \citenamefont {Peres}, \citenamefont {Novoselov},\ and\
  \citenamefont {Geim}}]{CastroNeto2009b}%
  \BibitemOpen
  \bibfield  {author} {\bibinfo {author} {\bibfnamefont {A.~H.}\ \bibnamefont
  {Castro~Neto}}, \bibinfo {author} {\bibfnamefont {F.}~\bibnamefont {Guinea}},
  \bibinfo {author} {\bibfnamefont {N.~M.~R.}\ \bibnamefont {Peres}}, \bibinfo
  {author} {\bibfnamefont {K.~S.}\ \bibnamefont {Novoselov}}, \ and\ \bibinfo
  {author} {\bibfnamefont {A.~K.}\ \bibnamefont {Geim}},\ }\bibfield  {title}
  {\enquote {\bibinfo {title} {The electronic properties of graphene},}\ }\href
  {\doibase 10.1103/RevModPhys.81.109} {\bibfield  {journal} {\bibinfo
  {journal} {Rev. Mod. Phys.}\ }\textbf {\bibinfo {volume} {81}},\ \bibinfo
  {pages} {109} (\bibinfo {year} {2009}{\natexlab{a}})}\BibitemShut {NoStop}%
\bibitem [{\citenamefont {Das~Sarma}\ \emph {et~al.}(2011)\citenamefont
  {Das~Sarma}, \citenamefont {Adam}, \citenamefont {Hwang},\ and\ \citenamefont
  {Rossi}}]{DasSarma2011}%
  \BibitemOpen
  \bibfield  {author} {\bibinfo {author} {\bibfnamefont {S.}~\bibnamefont
  {Das~Sarma}}, \bibinfo {author} {\bibfnamefont {S.}~\bibnamefont {Adam}},
  \bibinfo {author} {\bibfnamefont {E.~H.}\ \bibnamefont {Hwang}}, \ and\
  \bibinfo {author} {\bibfnamefont {E.}~\bibnamefont {Rossi}},\ }\bibfield
  {title} {\enquote {\bibinfo {title} {Electronic transport in two-dimensional
  graphene},}\ }\href@noop {} {\bibfield  {journal} {\bibinfo  {journal} {Rev.
  Mod. Phys.}\ }\textbf {\bibinfo {volume} {83}},\ \bibinfo {pages} {407}
  (\bibinfo {year} {2011})}\BibitemShut {NoStop}%
\bibitem [{\citenamefont {Zutic}\ \emph {et~al.}(2004)\citenamefont {Zutic},
  \citenamefont {Fabian},\ and\ \citenamefont {Das~Sarma}}]{Zutic2004}%
  \BibitemOpen
  \bibfield  {author} {\bibinfo {author} {\bibfnamefont {I.}~\bibnamefont
  {Zutic}}, \bibinfo {author} {\bibfnamefont {J.}~\bibnamefont {Fabian}}, \
  and\ \bibinfo {author} {\bibfnamefont {S.}~\bibnamefont {Das~Sarma}},\
  }\bibfield  {title} {\enquote {\bibinfo {title} {Spintronics: Fundamentals
  and applications},}\ }\href {\doibase 10.1103/RevModPhys.76.323} {\bibfield
  {journal} {\bibinfo  {journal} {Rev. Mod. Phys.}\ }\textbf {\bibinfo {volume}
  {76}},\ \bibinfo {pages} {323} (\bibinfo {year} {2004})}\BibitemShut
  {NoStop}%
\bibitem [{\citenamefont {Roche}\ and\ \citenamefont
  {Valenzuela}(2014)}]{Roche2014}%
  \BibitemOpen
  \bibfield  {author} {\bibinfo {author} {\bibfnamefont {S.}~\bibnamefont
  {Roche}}\ and\ \bibinfo {author} {\bibfnamefont {S.~O.}\ \bibnamefont
  {Valenzuela}},\ }\bibfield  {title} {\enquote {\bibinfo {title} {Graphene
  spintronics: puzzling controversies and challenges for spin manipulation},}\
  }\href {http://stacks.iop.org/0022-3727/47/i=9/a=094011} {\bibfield
  {journal} {\bibinfo  {journal} {Journal of Physics D: Applied Physics}\
  }\textbf {\bibinfo {volume} {47}},\ \bibinfo {pages} {094011} (\bibinfo
  {year} {2014})}\BibitemShut {NoStop}%
\bibitem [{\citenamefont {Huertas-Hernando}\ \emph {et~al.}(2006)\citenamefont
  {Huertas-Hernando}, \citenamefont {Guinea},\ and\ \citenamefont
  {Brataas}}]{Huertas-Hernando2006}%
  \BibitemOpen
  \bibfield  {author} {\bibinfo {author} {\bibfnamefont {D.}~\bibnamefont
  {Huertas-Hernando}}, \bibinfo {author} {\bibfnamefont {F.}~\bibnamefont
  {Guinea}}, \ and\ \bibinfo {author} {\bibfnamefont {A.}~\bibnamefont
  {Brataas}},\ }\bibfield  {title} {\enquote {\bibinfo {title} {Spin-orbit
  coupling in curved graphene, fullerenes, nanotubes, and nanotube caps},}\
  }\href {\doibase 10.1103/PhysRevB.74.155426} {\bibfield  {journal} {\bibinfo
  {journal} {Phys. Rev. B}\ }\textbf {\bibinfo {volume} {74}},\ \bibinfo
  {pages} {155426} (\bibinfo {year} {2006})}\BibitemShut {NoStop}%
\bibitem [{\citenamefont {Min}\ \emph {et~al.}(2006)\citenamefont {Min},
  \citenamefont {Hill}, \citenamefont {Sinitsyn}, \citenamefont {Sahu},
  \citenamefont {Kleinman},\ and\ \citenamefont {MacDonald}}]{Min2006}%
  \BibitemOpen
  \bibfield  {author} {\bibinfo {author} {\bibfnamefont {H.}~\bibnamefont
  {Min}}, \bibinfo {author} {\bibfnamefont {J.~E.}\ \bibnamefont {Hill}},
  \bibinfo {author} {\bibfnamefont {N.~A.}\ \bibnamefont {Sinitsyn}}, \bibinfo
  {author} {\bibfnamefont {B.~R.}\ \bibnamefont {Sahu}}, \bibinfo {author}
  {\bibfnamefont {L.}~\bibnamefont {Kleinman}}, \ and\ \bibinfo {author}
  {\bibfnamefont {A.~H.}\ \bibnamefont {MacDonald}},\ }\bibfield  {title}
  {\enquote {\bibinfo {title} {Intrinsic and rashba spin-orbit interactions in
  graphene sheets},}\ }\href {\doibase 10.1103/PhysRevB.74.165310} {\bibfield
  {journal} {\bibinfo  {journal} {Phys. Rev. B}\ }\textbf {\bibinfo {volume}
  {74}},\ \bibinfo {pages} {165310} (\bibinfo {year} {2006})}\BibitemShut
  {NoStop}%
\bibitem [{\citenamefont {Huertas-Hernando}\ \emph {et~al.}(2009)\citenamefont
  {Huertas-Hernando}, \citenamefont {Guinea},\ and\ \citenamefont
  {Brataas}}]{Huertas-Hernando2009}%
  \BibitemOpen
  \bibfield  {author} {\bibinfo {author} {\bibfnamefont {D.}~\bibnamefont
  {Huertas-Hernando}}, \bibinfo {author} {\bibfnamefont {F.}~\bibnamefont
  {Guinea}}, \ and\ \bibinfo {author} {\bibfnamefont {A.}~\bibnamefont
  {Brataas}},\ }\bibfield  {title} {\enquote {\bibinfo {title}
  {Spin-orbit-mediated spin relaxation in graphene},}\ }\href@noop {}
  {\bibfield  {journal} {\bibinfo  {journal} {Phys. Rev. Lett.}\ }\textbf
  {\bibinfo {volume} {103}},\ \bibinfo {pages} {146801} (\bibinfo {year}
  {2009})}\BibitemShut {NoStop}%
\bibitem [{\citenamefont {Ertler}\ \emph {et~al.}(2009)\citenamefont {Ertler},
  \citenamefont {Konschuh}, \citenamefont {Gmitra},\ and\ \citenamefont
  {Fabian}}]{Ertler2009}%
  \BibitemOpen
  \bibfield  {author} {\bibinfo {author} {\bibfnamefont {C.}~\bibnamefont
  {Ertler}}, \bibinfo {author} {\bibfnamefont {S.}~\bibnamefont {Konschuh}},
  \bibinfo {author} {\bibfnamefont {M.}~\bibnamefont {Gmitra}}, \ and\ \bibinfo
  {author} {\bibfnamefont {J.}~\bibnamefont {Fabian}},\ }\bibfield  {title}
  {\enquote {\bibinfo {title} {Electron spin relaxation in graphene: The role
  of the substrate},}\ }\href {\doibase 10.1103/PhysRevB.80.041405} {\bibfield
  {journal} {\bibinfo  {journal} {Phys. Rev. B}\ }\textbf {\bibinfo {volume}
  {80}},\ \bibinfo {pages} {041405} (\bibinfo {year} {2009})}\BibitemShut
  {NoStop}%
\bibitem [{\citenamefont {Castro~Neto}\ and\ \citenamefont
  {Guinea}(2009)}]{CastroNeto2009a}%
  \BibitemOpen
  \bibfield  {author} {\bibinfo {author} {\bibfnamefont {A.~H.}\ \bibnamefont
  {Castro~Neto}}\ and\ \bibinfo {author} {\bibfnamefont {F.}~\bibnamefont
  {Guinea}},\ }\bibfield  {title} {\enquote {\bibinfo {title} {Impurity-induced
  spin-orbit coupling in graphene},}\ }\href@noop {} {\bibfield  {journal}
  {\bibinfo  {journal} {Phys. Rev. Lett.}\ }\textbf {\bibinfo {volume} {103}},\
  \bibinfo {pages} {026804} (\bibinfo {year} {2009})}\BibitemShut {NoStop}%
\bibitem [{\citenamefont {Tombros}\ \emph {et~al.}(2007)\citenamefont
  {Tombros}, \citenamefont {Jozsa}, \citenamefont {Popinciuc}, \citenamefont
  {Jonkman},\ and\ \citenamefont {van Wees}}]{Tombros2007}%
  \BibitemOpen
  \bibfield  {author} {\bibinfo {author} {\bibfnamefont {N.}~\bibnamefont
  {Tombros}}, \bibinfo {author} {\bibfnamefont {C.}~\bibnamefont {Jozsa}},
  \bibinfo {author} {\bibfnamefont {M.}~\bibnamefont {Popinciuc}}, \bibinfo
  {author} {\bibfnamefont {H.~T.}\ \bibnamefont {Jonkman}}, \ and\ \bibinfo
  {author} {\bibfnamefont {B.~J.}\ \bibnamefont {van Wees}},\ }\bibfield
  {title} {\enquote {\bibinfo {title} {Electronic spin transport and spin
  precession in single graphene layers at room temperature},}\ }\href {\doibase
  10.1038/nature06037, Letter} {\bibfield  {journal} {\bibinfo  {journal}
  {Nature}\ }\textbf {\bibinfo {volume} {448}},\ \bibinfo {pages} {571}
  (\bibinfo {year} {2007})}\BibitemShut {NoStop}%
\bibitem [{\citenamefont {Han}\ \emph {et~al.}(2010)\citenamefont {Han},
  \citenamefont {Pi}, \citenamefont {McCreary}, \citenamefont {Li},
  \citenamefont {Wong}, \citenamefont {Swartz},\ and\ \citenamefont
  {Kawakami}}]{Han2010a}%
  \BibitemOpen
  \bibfield  {author} {\bibinfo {author} {\bibfnamefont {W.}~\bibnamefont
  {Han}}, \bibinfo {author} {\bibfnamefont {K.}~\bibnamefont {Pi}}, \bibinfo
  {author} {\bibfnamefont {K.~M.}\ \bibnamefont {McCreary}}, \bibinfo {author}
  {\bibfnamefont {Y.}~\bibnamefont {Li}}, \bibinfo {author} {\bibfnamefont
  {J.~J.~I.}\ \bibnamefont {Wong}}, \bibinfo {author} {\bibfnamefont {A.~G.}\
  \bibnamefont {Swartz}}, \ and\ \bibinfo {author} {\bibfnamefont {R.~K.}\
  \bibnamefont {Kawakami}},\ }\bibfield  {title} {\enquote {\bibinfo {title}
  {Tunneling spin injection into single layer graphene},}\ }\href {\doibase
  10.1103/PhysRevLett.105.167202} {\bibfield  {journal} {\bibinfo  {journal}
  {Phys. Rev. Lett.}\ }\textbf {\bibinfo {volume} {105}},\ \bibinfo {pages}
  {167202} (\bibinfo {year} {2010})}\BibitemShut {NoStop}%
\bibitem [{\citenamefont {Han}\ and\ \citenamefont {Kawakami}(2011)}]{Han2011}%
  \BibitemOpen
  \bibfield  {author} {\bibinfo {author} {\bibfnamefont {W.}~\bibnamefont
  {Han}}\ and\ \bibinfo {author} {\bibfnamefont {R.~K.}\ \bibnamefont
  {Kawakami}},\ }\bibfield  {title} {\enquote {\bibinfo {title} {Spin
  relaxation in single-layer and bilayer graphene},}\ }\href {\doibase
  10.1103/PhysRevLett.107.047207} {\bibfield  {journal} {\bibinfo  {journal}
  {Phys. Rev. Lett.}\ }\textbf {\bibinfo {volume} {107}},\ \bibinfo {pages}
  {047207} (\bibinfo {year} {2011})}\BibitemShut {NoStop}%
\bibitem [{\citenamefont {Zomer}\ \emph {et~al.}(2012)\citenamefont {Zomer},
  \citenamefont {Guimar\~aes}, \citenamefont {Tombros},\ and\ \citenamefont
  {van Wees}}]{Zomer2012}%
  \BibitemOpen
  \bibfield  {author} {\bibinfo {author} {\bibfnamefont {P.~J.}\ \bibnamefont
  {Zomer}}, \bibinfo {author} {\bibfnamefont {M.~H.~D.}\ \bibnamefont
  {Guimar\~aes}}, \bibinfo {author} {\bibfnamefont {N.}~\bibnamefont
  {Tombros}}, \ and\ \bibinfo {author} {\bibfnamefont {B.~J.}\ \bibnamefont
  {van Wees}},\ }\bibfield  {title} {\enquote {\bibinfo {title} {Long-distance
  spin transport in high-mobility graphene on hexagonal boron nitride},}\
  }\href {\doibase 10.1103/PhysRevB.86.161416} {\bibfield  {journal} {\bibinfo
  {journal} {Phys. Rev. B}\ }\textbf {\bibinfo {volume} {86}},\ \bibinfo
  {pages} {161416} (\bibinfo {year} {2012})}\BibitemShut {NoStop}%
\bibitem [{\citenamefont {Maassen}\ \emph {et~al.}(2012)\citenamefont
  {Maassen}, \citenamefont {Vera-Marun}, \citenamefont {Guimar\~aes},\ and\
  \citenamefont {van Wees}}]{Maassen2012}%
  \BibitemOpen
  \bibfield  {author} {\bibinfo {author} {\bibfnamefont {T.}~\bibnamefont
  {Maassen}}, \bibinfo {author} {\bibfnamefont {I.~J.}\ \bibnamefont
  {Vera-Marun}}, \bibinfo {author} {\bibfnamefont {M.~H.~D.}\ \bibnamefont
  {Guimar\~aes}}, \ and\ \bibinfo {author} {\bibfnamefont {B.~J.}\ \bibnamefont
  {van Wees}},\ }\bibfield  {title} {\enquote {\bibinfo {title}
  {Contact-induced spin relaxation in hanle spin precession measurements},}\
  }\href {\doibase 10.1103/PhysRevB.86.235408} {\bibfield  {journal} {\bibinfo
  {journal} {Phys. Rev. B}\ }\textbf {\bibinfo {volume} {86}},\ \bibinfo
  {pages} {235408} (\bibinfo {year} {2012})}\BibitemShut {NoStop}%
\bibitem [{\citenamefont {Lundeberg}\ \emph {et~al.}(2013)\citenamefont
  {Lundeberg}, \citenamefont {Yang}, \citenamefont {Renard},\ and\
  \citenamefont {Folk}}]{Lundeberg2013}%
  \BibitemOpen
  \bibfield  {author} {\bibinfo {author} {\bibfnamefont {M.}~\bibnamefont
  {Lundeberg}}, \bibinfo {author} {\bibfnamefont {R.}~\bibnamefont {Yang}},
  \bibinfo {author} {\bibfnamefont {J.}~\bibnamefont {Renard}}, \ and\ \bibinfo
  {author} {\bibfnamefont {J.}~\bibnamefont {Folk}},\ }\bibfield  {title}
  {\enquote {\bibinfo {title} {Defect-mediated spin relaxation and dephasing in
  graphene},}\ }\href@noop {} {\bibfield  {journal} {\bibinfo  {journal} {Phys.
  Rev. Lett.}\ }\textbf {\bibinfo {volume} {110}},\ \bibinfo {pages} {156601}
  (\bibinfo {year} {2013})}\BibitemShut {NoStop}%
\bibitem [{\citenamefont {{Idzuchi}}\ \emph {et~al.}(2014)\citenamefont
  {{Idzuchi}}, \citenamefont {{Fert}},\ and\ \citenamefont
  {{Otani}}}]{Idzuchi2014}%
  \BibitemOpen
  \bibfield  {author} {\bibinfo {author} {\bibfnamefont {H.}~\bibnamefont
  {{Idzuchi}}}, \bibinfo {author} {\bibfnamefont {A.}~\bibnamefont {{Fert}}}, \
  and\ \bibinfo {author} {\bibfnamefont {Y.}~\bibnamefont {{Otani}}},\
  }\bibfield  {title} {\enquote {\bibinfo {title} {Revisiting the measurement
  of the spin relaxation time in graphene-based devices},}\ }\href@noop {}
  {\bibfield  {journal} {\bibinfo  {journal} {ArXiv e-prints}\ } (\bibinfo
  {year} {2014})},\ \Eprint {http://arxiv.org/abs/1411.2949} {arXiv:1411.2949
  [cond-mat.mes-hall]} \BibitemShut {NoStop}%
\bibitem [{\citenamefont {Yazyev}\ and\ \citenamefont
  {Helm}(2007)}]{Yazyev2007}%
  \BibitemOpen
  \bibfield  {author} {\bibinfo {author} {\bibfnamefont {O.~V.}\ \bibnamefont
  {Yazyev}}\ and\ \bibinfo {author} {\bibfnamefont {L.}~\bibnamefont {Helm}},\
  }\bibfield  {title} {\enquote {\bibinfo {title} {Defect-induced magnetism in
  graphene},}\ }\href@noop {} {\bibfield  {journal} {\bibinfo  {journal} {Phys.
  Rev. B}\ }\textbf {\bibinfo {volume} {75}},\ \bibinfo {pages} {125408}
  (\bibinfo {year} {2007})}\BibitemShut {NoStop}%
\bibitem [{\citenamefont {Palacios}\ \emph {et~al.}(2008)\citenamefont
  {Palacios}, \citenamefont {Fern{\'a}ndez-Rossier},\ and\ \citenamefont
  {Brey}}]{Palacios2008}%
  \BibitemOpen
  \bibfield  {author} {\bibinfo {author} {\bibfnamefont {J.}~\bibnamefont
  {Palacios}}, \bibinfo {author} {\bibfnamefont {J.}~\bibnamefont
  {Fern{\'a}ndez-Rossier}}, \ and\ \bibinfo {author} {\bibfnamefont
  {L.}~\bibnamefont {Brey}},\ }\bibfield  {title} {\enquote {\bibinfo {title}
  {Vacancy-induced magnetism in graphene and graphene ribbons},}\ }\href@noop
  {} {\bibfield  {journal} {\bibinfo  {journal} {Phys. Rev. B}\ }\textbf
  {\bibinfo {volume} {77}},\ \bibinfo {pages} {195428} (\bibinfo {year}
  {2008})}\BibitemShut {NoStop}%
\bibitem [{\citenamefont {Yazyev}(2008)}]{Yazyev2008}%
  \BibitemOpen
  \bibfield  {author} {\bibinfo {author} {\bibfnamefont {O.~V.}\ \bibnamefont
  {Yazyev}},\ }\bibfield  {title} {\enquote {\bibinfo {title} {Magnetism in
  disordered graphene and irradiated graphite},}\ }\href@noop {} {\bibfield
  {journal} {\bibinfo  {journal} {Phys. Rev. Lett.}\ }\textbf {\bibinfo
  {volume} {101}},\ \bibinfo {pages} {037203} (\bibinfo {year}
  {2008})}\BibitemShut {NoStop}%
\bibitem [{\citenamefont {Balog}\ \emph {et~al.}(2009)\citenamefont {Balog},
  \citenamefont {J{\o}rgensen}, \citenamefont {Wells}, \citenamefont
  {L{\ae}gsgaard}, \citenamefont {Hofmann}, \citenamefont {Besenbacher},\ and\
  \citenamefont {Hornek{\ae}r}}]{Balog2009}%
  \BibitemOpen
  \bibfield  {author} {\bibinfo {author} {\bibfnamefont {R.}~\bibnamefont
  {Balog}}, \bibinfo {author} {\bibfnamefont {B.}~\bibnamefont {J{\o}rgensen}},
  \bibinfo {author} {\bibfnamefont {J.}~\bibnamefont {Wells}}, \bibinfo
  {author} {\bibfnamefont {E.}~\bibnamefont {L{\ae}gsgaard}}, \bibinfo {author}
  {\bibfnamefont {P.}~\bibnamefont {Hofmann}}, \bibinfo {author} {\bibfnamefont
  {F.}~\bibnamefont {Besenbacher}}, \ and\ \bibinfo {author} {\bibfnamefont
  {L.}~\bibnamefont {Hornek{\ae}r}},\ }\bibfield  {title} {\enquote {\bibinfo
  {title} {Atomic hydrogen adsorbate structures on graphene},}\ }\href@noop {}
  {\bibfield  {journal} {\bibinfo  {journal} {J. Am. Chem. Soc.}\ }\textbf
  {\bibinfo {volume} {131}},\ \bibinfo {pages} {8744} (\bibinfo {year}
  {2009})}\BibitemShut {NoStop}%
\bibitem [{\citenamefont {Yazyev}(2010)}]{Yazyev2010}%
  \BibitemOpen
  \bibfield  {author} {\bibinfo {author} {\bibfnamefont {O.~V.}\ \bibnamefont
  {Yazyev}},\ }\bibfield  {title} {\enquote {\bibinfo {title} {Emergence of
  magnetism in graphene materials and nanostructures},}\ }\href@noop {}
  {\bibfield  {journal} {\bibinfo  {journal} {Reports on Progress in Physics}\
  }\textbf {\bibinfo {volume} {73}},\ \bibinfo {pages} {056501} (\bibinfo
  {year} {2010})}\BibitemShut {NoStop}%
\bibitem [{\citenamefont {Haase}\ \emph {et~al.}(2011)\citenamefont {Haase},
  \citenamefont {Fuchs}, \citenamefont {Pruschke}, \citenamefont {Ochoa},\ and\
  \citenamefont {Guinea}}]{Haase2011}%
  \BibitemOpen
  \bibfield  {author} {\bibinfo {author} {\bibfnamefont {P.}~\bibnamefont
  {Haase}}, \bibinfo {author} {\bibfnamefont {S.}~\bibnamefont {Fuchs}},
  \bibinfo {author} {\bibfnamefont {T.}~\bibnamefont {Pruschke}}, \bibinfo
  {author} {\bibfnamefont {H.}~\bibnamefont {Ochoa}}, \ and\ \bibinfo {author}
  {\bibfnamefont {F.}~\bibnamefont {Guinea}},\ }\bibfield  {title} {\enquote
  {\bibinfo {title} {Magnetic moments and kondo effect near vacancies and
  resonant scatterers in graphene},}\ }\href@noop {} {\bibfield  {journal}
  {\bibinfo  {journal} {Phys. Rev. B}\ }\textbf {\bibinfo {volume} {83}},\
  \bibinfo {pages} {241408} (\bibinfo {year} {2011})}\BibitemShut {NoStop}%
\bibitem [{\citenamefont {Uchoa}\ \emph {et~al.}(2008)\citenamefont {Uchoa},
  \citenamefont {Kotov}, \citenamefont {Peres},\ and\ \citenamefont
  {Castro~Neto}}]{Uchoa2008}%
  \BibitemOpen
  \bibfield  {author} {\bibinfo {author} {\bibfnamefont {B.}~\bibnamefont
  {Uchoa}}, \bibinfo {author} {\bibfnamefont {V.}~\bibnamefont {Kotov}},
  \bibinfo {author} {\bibfnamefont {N.~M.~R.}\ \bibnamefont {Peres}}, \ and\
  \bibinfo {author} {\bibfnamefont {A.~H.}\ \bibnamefont {Castro~Neto}},\
  }\bibfield  {title} {\enquote {\bibinfo {title} {Localized magnetic states in
  graphene},}\ }\href@noop {} {\bibfield  {journal} {\bibinfo  {journal} {Phys.
  Rev. Lett.}\ }\textbf {\bibinfo {volume} {101}},\ \bibinfo {pages} {026805}
  (\bibinfo {year} {2008})}\BibitemShut {NoStop}%
\bibitem [{\citenamefont {Lehtinen}\ \emph {et~al.}(2003)\citenamefont
  {Lehtinen}, \citenamefont {Foster}, \citenamefont {Ayuela}, \citenamefont
  {Krasheninnikov}, \citenamefont {Nordlund},\ and\ \citenamefont
  {Nieminen}}]{Lehtinen2003}%
  \BibitemOpen
  \bibfield  {author} {\bibinfo {author} {\bibfnamefont {P.}~\bibnamefont
  {Lehtinen}}, \bibinfo {author} {\bibfnamefont {A.}~\bibnamefont {Foster}},
  \bibinfo {author} {\bibfnamefont {A.}~\bibnamefont {Ayuela}}, \bibinfo
  {author} {\bibfnamefont {A.}~\bibnamefont {Krasheninnikov}}, \bibinfo
  {author} {\bibfnamefont {K.}~\bibnamefont {Nordlund}}, \ and\ \bibinfo
  {author} {\bibfnamefont {R.}~\bibnamefont {Nieminen}},\ }\bibfield  {title}
  {\enquote {\bibinfo {title} {Magnetic properties and diffusion of adatoms on
  a graphene sheet},}\ }\href@noop {} {\bibfield  {journal} {\bibinfo
  {journal} {Phys. Rev. Lett.}\ }\textbf {\bibinfo {volume} {91}},\ \bibinfo
  {pages} {017202} (\bibinfo {year} {2003})}\BibitemShut {NoStop}%
\bibitem [{\citenamefont {Duplock}\ \emph {et~al.}(2004)\citenamefont
  {Duplock}, \citenamefont {Scheffler},\ and\ \citenamefont
  {Lindan}}]{Duplock2004}%
  \BibitemOpen
  \bibfield  {author} {\bibinfo {author} {\bibfnamefont {E.}~\bibnamefont
  {Duplock}}, \bibinfo {author} {\bibfnamefont {M.}~\bibnamefont {Scheffler}},
  \ and\ \bibinfo {author} {\bibfnamefont {P.}~\bibnamefont {Lindan}},\
  }\bibfield  {title} {\enquote {\bibinfo {title} {Hallmark of perfect
  graphene},}\ }\href@noop {} {\bibfield  {journal} {\bibinfo  {journal} {Phys.
  Rev. Lett.}\ }\textbf {\bibinfo {volume} {92}},\ \bibinfo {pages} {225502}
  (\bibinfo {year} {2004})}\BibitemShut {NoStop}%
\bibitem [{\citenamefont {Meyer}\ \emph {et~al.}(2008)\citenamefont {Meyer},
  \citenamefont {Girit}, \citenamefont {Crommie},\ and\ \citenamefont
  {Zettl}}]{Meyer2008}%
  \BibitemOpen
  \bibfield  {author} {\bibinfo {author} {\bibfnamefont {J.~C.}\ \bibnamefont
  {Meyer}}, \bibinfo {author} {\bibfnamefont {C.~O.}\ \bibnamefont {Girit}},
  \bibinfo {author} {\bibfnamefont {M.~F.}\ \bibnamefont {Crommie}}, \ and\
  \bibinfo {author} {\bibfnamefont {A.}~\bibnamefont {Zettl}},\ }\bibfield
  {title} {\enquote {\bibinfo {title} {Imaging and dynamics of light atoms and
  molecules on graphene},}\ }\href@noop {} {\bibfield  {journal} {\bibinfo
  {journal} {Nature}\ }\textbf {\bibinfo {volume} {454}},\ \bibinfo {pages}
  {319} (\bibinfo {year} {2008})}\BibitemShut {NoStop}%
\bibitem [{\citenamefont {Chan}\ \emph {et~al.}(2008)\citenamefont {Chan},
  \citenamefont {Neaton},\ and\ \citenamefont {Cohen}}]{Chan2008}%
  \BibitemOpen
  \bibfield  {author} {\bibinfo {author} {\bibfnamefont {K.}~\bibnamefont
  {Chan}}, \bibinfo {author} {\bibfnamefont {J.~B.}\ \bibnamefont {Neaton}}, \
  and\ \bibinfo {author} {\bibfnamefont {M.~L.}\ \bibnamefont {Cohen}},\
  }\bibfield  {title} {\enquote {\bibinfo {title} {First-principles study of
  metal adatom adsorption on graphene},}\ }\href@noop {} {\bibfield  {journal}
  {\bibinfo  {journal} {Phys. Rev. B}\ }\textbf {\bibinfo {volume} {77}},\
  \bibinfo {pages} {235430} (\bibinfo {year} {2008})}\BibitemShut {NoStop}%
\bibitem [{\citenamefont {Castro~Neto}\ \emph
  {et~al.}(2009{\natexlab{b}})\citenamefont {Castro~Neto}, \citenamefont
  {Kotov}, \citenamefont {Nilsson}, \citenamefont {Pereira}, \citenamefont
  {Peres},\ and\ \citenamefont {Uchoa}}]{CastroNeto2009c}%
  \BibitemOpen
  \bibfield  {author} {\bibinfo {author} {\bibfnamefont {A.~H.}\ \bibnamefont
  {Castro~Neto}}, \bibinfo {author} {\bibfnamefont {V.~N.}\ \bibnamefont
  {Kotov}}, \bibinfo {author} {\bibfnamefont {J.}~\bibnamefont {Nilsson}},
  \bibinfo {author} {\bibfnamefont {V.~M.}\ \bibnamefont {Pereira}}, \bibinfo
  {author} {\bibfnamefont {N.~M.~R.}\ \bibnamefont {Peres}}, \ and\ \bibinfo
  {author} {\bibfnamefont {B.}~\bibnamefont {Uchoa}},\ }\bibfield  {title}
  {\enquote {\bibinfo {title} {Adatoms in graphene},}\ }\href@noop {}
  {\bibfield  {journal} {\bibinfo  {journal} {Solid State Comm.}\ }\textbf
  {\bibinfo {volume} {149}},\ \bibinfo {pages} {1094} (\bibinfo {year}
  {2009}{\natexlab{b}})}\BibitemShut {NoStop}%
\bibitem [{\citenamefont {Boukhvalov}\ \emph {et~al.}(2008)\citenamefont
  {Boukhvalov}, \citenamefont {Katsnelson},\ and\ \citenamefont
  {Lichtenstein}}]{Boukhvalov2008}%
  \BibitemOpen
  \bibfield  {author} {\bibinfo {author} {\bibfnamefont {D.}~\bibnamefont
  {Boukhvalov}}, \bibinfo {author} {\bibfnamefont {M.}~\bibnamefont
  {Katsnelson}}, \ and\ \bibinfo {author} {\bibfnamefont {A.}~\bibnamefont
  {Lichtenstein}},\ }\bibfield  {title} {\enquote {\bibinfo {title} {Hydrogen
  on graphene: Electronic structure, total energy, structural distortions and
  magnetism from first-principles calculations},}\ }\href@noop {} {\bibfield
  {journal} {\bibinfo  {journal} {Phys. Rev. B}\ }\textbf {\bibinfo {volume}
  {77}},\ \bibinfo {pages} {035427} (\bibinfo {year} {2008})}\BibitemShut
  {NoStop}%
\bibitem [{\citenamefont {Boukhvalov}\ and\ \citenamefont
  {Katsnelson}(2009)}]{Boukhvalov2009}%
  \BibitemOpen
  \bibfield  {author} {\bibinfo {author} {\bibfnamefont {D.~W.}\ \bibnamefont
  {Boukhvalov}}\ and\ \bibinfo {author} {\bibfnamefont {M.~I.}\ \bibnamefont
  {Katsnelson}},\ }\bibfield  {title} {\enquote {\bibinfo {title} {Chemical
  functionalization of graphene},}\ }\href@noop {} {\bibfield  {journal}
  {\bibinfo  {journal} {J. Phys.: Condens. Matter}\ }\textbf {\bibinfo {volume}
  {21}},\ \bibinfo {pages} {344205} (\bibinfo {year} {2009})}\BibitemShut
  {NoStop}%
\bibitem [{\citenamefont {Cornaglia}\ \emph {et~al.}(2009)\citenamefont
  {Cornaglia}, \citenamefont {Usaj},\ and\ \citenamefont
  {Balseiro}}]{Cornaglia2009}%
  \BibitemOpen
  \bibfield  {author} {\bibinfo {author} {\bibfnamefont {P.~S.}\ \bibnamefont
  {Cornaglia}}, \bibinfo {author} {\bibfnamefont {G.}~\bibnamefont {Usaj}}, \
  and\ \bibinfo {author} {\bibfnamefont {C.~A.}\ \bibnamefont {Balseiro}},\
  }\bibfield  {title} {\enquote {\bibinfo {title} {Localized spins on
  graphene},}\ }\href {\doibase 10.1103/PhysRevLett.102.046801} {\bibfield
  {journal} {\bibinfo  {journal} {Phys. Rev. Lett.}\ }\textbf {\bibinfo
  {volume} {102}},\ \bibinfo {pages} {046801} (\bibinfo {year}
  {2009})}\BibitemShut {NoStop}%
\bibitem [{\citenamefont {Wehling}\ \emph {et~al.}(2009)\citenamefont
  {Wehling}, \citenamefont {Katsnelson},\ and\ \citenamefont
  {Lichtenstein}}]{Wehling2009}%
  \BibitemOpen
  \bibfield  {author} {\bibinfo {author} {\bibfnamefont {T.}~\bibnamefont
  {Wehling}}, \bibinfo {author} {\bibfnamefont {M.}~\bibnamefont {Katsnelson}},
  \ and\ \bibinfo {author} {\bibfnamefont {A.}~\bibnamefont {Lichtenstein}},\
  }\bibfield  {title} {\enquote {\bibinfo {title} {Impurities on graphene:
  Midgap states and migration barriers},}\ }\href@noop {} {\bibfield  {journal}
  {\bibinfo  {journal} {Phys. Rev. B}\ }\textbf {\bibinfo {volume} {80}},\
  \bibinfo {pages} {085428} (\bibinfo {year} {2009})}\BibitemShut {NoStop}%
\bibitem [{\citenamefont {Wehling}\ \emph
  {et~al.}(2010{\natexlab{a}})\citenamefont {Wehling}, \citenamefont
  {Balatsky}, \citenamefont {Katsnelson}, \citenamefont {Lichtenstein},\ and\
  \citenamefont {Rosch}}]{Wehling2010}%
  \BibitemOpen
  \bibfield  {author} {\bibinfo {author} {\bibfnamefont {T.~O.}\ \bibnamefont
  {Wehling}}, \bibinfo {author} {\bibfnamefont {A.~V.}\ \bibnamefont
  {Balatsky}}, \bibinfo {author} {\bibfnamefont {M.~I.}\ \bibnamefont
  {Katsnelson}}, \bibinfo {author} {\bibfnamefont {A.~I.}\ \bibnamefont
  {Lichtenstein}}, \ and\ \bibinfo {author} {\bibfnamefont {A.}~\bibnamefont
  {Rosch}},\ }\bibfield  {title} {\enquote {\bibinfo {title} {Orbitally
  controlled kondo effect of co adatoms on graphene},}\ }\href@noop {}
  {\bibfield  {journal} {\bibinfo  {journal} {Phys. Rev. B}\ }\textbf {\bibinfo
  {volume} {81}},\ \bibinfo {pages} {115427} (\bibinfo {year}
  {2010}{\natexlab{a}})}\BibitemShut {NoStop}%
\bibitem [{\citenamefont {Wehling}\ \emph
  {et~al.}(2010{\natexlab{b}})\citenamefont {Wehling}, \citenamefont {Dahal},
  \citenamefont {Lichtenstein}, \citenamefont {Katsnelson}, \citenamefont
  {Manoharan},\ and\ \citenamefont {Balatsky}}]{Wehling2010a}%
  \BibitemOpen
  \bibfield  {author} {\bibinfo {author} {\bibfnamefont {T.~O.}\ \bibnamefont
  {Wehling}}, \bibinfo {author} {\bibfnamefont {H.~P.}\ \bibnamefont {Dahal}},
  \bibinfo {author} {\bibfnamefont {A.~I.}\ \bibnamefont {Lichtenstein}},
  \bibinfo {author} {\bibfnamefont {M.~I.}\ \bibnamefont {Katsnelson}},
  \bibinfo {author} {\bibfnamefont {H.~C.}\ \bibnamefont {Manoharan}}, \ and\
  \bibinfo {author} {\bibfnamefont {A.~V.}\ \bibnamefont {Balatsky}},\
  }\bibfield  {title} {\enquote {\bibinfo {title} {Theory of fano resonances in
  graphene: The influence of orbital and structural symmetries on stm
  spectra},}\ }\href@noop {} {\bibfield  {journal} {\bibinfo  {journal} {Phys.
  Rev. B}\ }\textbf {\bibinfo {volume} {81}},\ \bibinfo {pages} {085413}
  (\bibinfo {year} {2010}{\natexlab{b}})}\BibitemShut {NoStop}%
\bibitem [{\citenamefont {Ao}\ and\ \citenamefont {Peeters}(2010)}]{Ao2010}%
  \BibitemOpen
  \bibfield  {author} {\bibinfo {author} {\bibfnamefont {Z.~M.}\ \bibnamefont
  {Ao}}\ and\ \bibinfo {author} {\bibfnamefont {F.~M.}\ \bibnamefont
  {Peeters}},\ }\bibfield  {title} {\enquote {\bibinfo {title} {Electric field:
  A catalyst for hydrogenation of graphene},}\ }\href@noop {} {\bibfield
  {journal} {\bibinfo  {journal} {Appl. Phys. Lett.}\ }\textbf {\bibinfo
  {volume} {96}},\ \bibinfo {pages} {253106} (\bibinfo {year}
  {2010})}\BibitemShut {NoStop}%
\bibitem [{\citenamefont {Chan}\ \emph {et~al.}(2011)\citenamefont {Chan},
  \citenamefont {Lee},\ and\ \citenamefont {Cohen}}]{Chan2011}%
  \BibitemOpen
  \bibfield  {author} {\bibinfo {author} {\bibfnamefont {K.}~\bibnamefont
  {Chan}}, \bibinfo {author} {\bibfnamefont {H.}~\bibnamefont {Lee}}, \ and\
  \bibinfo {author} {\bibfnamefont {M.~L.}\ \bibnamefont {Cohen}},\ }\bibfield
  {title} {\enquote {\bibinfo {title} {Possibility of transforming the
  electronic structure of one species of graphene adatoms into that of another
  by application of gate voltage: First-principles calculations},}\ }\href@noop
  {} {\bibfield  {journal} {\bibinfo  {journal} {Phys. Rev. B}\ }\textbf
  {\bibinfo {volume} {84}},\ \bibinfo {pages} {165419} (\bibinfo {year}
  {2011})}\BibitemShut {NoStop}%
\bibitem [{\citenamefont {Sofo}\ \emph {et~al.}(2012)\citenamefont {Sofo},
  \citenamefont {Usaj}, \citenamefont {Cornaglia}, \citenamefont {Suarez},
  \citenamefont {Hern{\'a}ndez-Nieves},\ and\ \citenamefont
  {Balseiro}}]{Sofo2012}%
  \BibitemOpen
  \bibfield  {author} {\bibinfo {author} {\bibfnamefont {J.~O.}\ \bibnamefont
  {Sofo}}, \bibinfo {author} {\bibfnamefont {G.}~\bibnamefont {Usaj}}, \bibinfo
  {author} {\bibfnamefont {P.~S.}\ \bibnamefont {Cornaglia}}, \bibinfo {author}
  {\bibfnamefont {A.~M.}\ \bibnamefont {Suarez}}, \bibinfo {author}
  {\bibfnamefont {A.~D.}\ \bibnamefont {Hern{\'a}ndez-Nieves}}, \ and\ \bibinfo
  {author} {\bibfnamefont {C.~A.}\ \bibnamefont {Balseiro}},\ }\bibfield
  {title} {\enquote {\bibinfo {title} {Magnetic structure of hydrogen-induced
  defects on graphene},}\ }\href {\doibase 10.1103/PhysRevB.85.115405}
  {\bibfield  {journal} {\bibinfo  {journal} {Phys. Rev. B}\ }\textbf {\bibinfo
  {volume} {85}},\ \bibinfo {pages} {115405} (\bibinfo {year}
  {2012})}\BibitemShut {NoStop}%
\bibitem [{\citenamefont {Kochan}\ \emph {et~al.}(2014)\citenamefont {Kochan},
  \citenamefont {Gmitra},\ and\ \citenamefont {Fabian}}]{Kochan2014}%
  \BibitemOpen
  \bibfield  {author} {\bibinfo {author} {\bibfnamefont {D.}~\bibnamefont
  {Kochan}}, \bibinfo {author} {\bibfnamefont {M.}~\bibnamefont {Gmitra}}, \
  and\ \bibinfo {author} {\bibfnamefont {J.}~\bibnamefont {Fabian}},\
  }\bibfield  {title} {\enquote {\bibinfo {title} {Spin relaxation mechanism in
  graphene: Resonant scattering by magnetic impurities},}\ }\href {\doibase
  10.1103/PhysRevLett.112.116602} {\bibfield  {journal} {\bibinfo  {journal}
  {Phys. Rev. Lett.}\ }\textbf {\bibinfo {volume} {112}},\ \bibinfo {pages}
  {116602} (\bibinfo {year} {2014})}\BibitemShut {NoStop}%
\bibitem [{\citenamefont {Song}\ and\ \citenamefont {Dery}(2013)}]{Song2013}%
  \BibitemOpen
  \bibfield  {author} {\bibinfo {author} {\bibfnamefont {Y.}~\bibnamefont
  {Song}}\ and\ \bibinfo {author} {\bibfnamefont {H.}~\bibnamefont {Dery}},\
  }\bibfield  {title} {\enquote {\bibinfo {title} {Transport theory of
  monolayer transition-metal dichalcogenides through symmetry},}\ }\href
  {\doibase 10.1103/PhysRevLett.111.026601} {\bibfield  {journal} {\bibinfo
  {journal} {Phys. Rev. Lett.}\ }\textbf {\bibinfo {volume} {111}},\ \bibinfo
  {pages} {026601} (\bibinfo {year} {2013})}\BibitemShut {NoStop}%
\bibitem [{\citenamefont {Tuan}\ \emph {et~al.}(2014)\citenamefont {Tuan},
  \citenamefont {Ortmann}, \citenamefont {Soriano}, \citenamefont
  {Valenzuela},\ and\ \citenamefont {Roche}}]{Tuan2014}%
  \BibitemOpen
  \bibfield  {author} {\bibinfo {author} {\bibfnamefont {D.~V.}\ \bibnamefont
  {Tuan}}, \bibinfo {author} {\bibfnamefont {F.}~\bibnamefont {Ortmann}},
  \bibinfo {author} {\bibfnamefont {D.}~\bibnamefont {Soriano}}, \bibinfo
  {author} {\bibfnamefont {S.~O.}\ \bibnamefont {Valenzuela}}, \ and\ \bibinfo
  {author} {\bibfnamefont {S.}~\bibnamefont {Roche}},\ }\bibfield  {title}
  {\enquote {\bibinfo {title} {Pseudospin-driven spin relaxation mechanism
  in~graphene},}\ }\href@noop {} {\bibfield  {journal} {\bibinfo  {journal}
  {Nat. Phys.}\ }\textbf {\bibinfo {volume} {10}},\ \bibinfo {pages} {857}
  (\bibinfo {year} {2014})}\BibitemShut {NoStop}%
\bibitem [{\citenamefont {Sofo}\ \emph {et~al.}(2011)\citenamefont {Sofo},
  \citenamefont {Suarez}, \citenamefont {Usaj}, \citenamefont {Cornaglia},
  \citenamefont {Hern{\'a}ndez-Nieves},\ and\ \citenamefont
  {Balseiro}}]{Sofo2011}%
  \BibitemOpen
  \bibfield  {author} {\bibinfo {author} {\bibfnamefont {J.~O.}\ \bibnamefont
  {Sofo}}, \bibinfo {author} {\bibfnamefont {A.~M.}\ \bibnamefont {Suarez}},
  \bibinfo {author} {\bibfnamefont {G.}~\bibnamefont {Usaj}}, \bibinfo {author}
  {\bibfnamefont {P.~S.}\ \bibnamefont {Cornaglia}}, \bibinfo {author}
  {\bibfnamefont {A.~D.}\ \bibnamefont {Hern{\'a}ndez-Nieves}}, \ and\ \bibinfo
  {author} {\bibfnamefont {C.~A.}\ \bibnamefont {Balseiro}},\ }\bibfield
  {title} {\enquote {\bibinfo {title} {Electrical control of the chemical
  bonding of fluorine on graphene},}\ }\href {\doibase
  10.1103/PhysRevB.83.081411} {\bibfield  {journal} {\bibinfo  {journal} {Phys.
  Rev. B}\ }\textbf {\bibinfo {volume} {83}},\ \bibinfo {pages} {081411}
  (\bibinfo {year} {2011})}\BibitemShut {NoStop}%
\bibitem [{\citenamefont {Guzm{\'a}n-Arellano}\ \emph
  {et~al.}(2014)\citenamefont {Guzm{\'a}n-Arellano}, \citenamefont
  {Hern{\'a}ndez-Nieves}, \citenamefont {Balseiro},\ and\ \citenamefont
  {Usaj}}]{Guzman2014}%
  \BibitemOpen
  \bibfield  {author} {\bibinfo {author} {\bibfnamefont {R.~M.}\ \bibnamefont
  {Guzm{\'a}n-Arellano}}, \bibinfo {author} {\bibfnamefont {A.~D.}\
  \bibnamefont {Hern{\'a}ndez-Nieves}}, \bibinfo {author} {\bibfnamefont
  {C.~A.}\ \bibnamefont {Balseiro}}, \ and\ \bibinfo {author} {\bibfnamefont
  {G.}~\bibnamefont {Usaj}},\ }\bibfield  {title} {\enquote {\bibinfo {title}
  {Diffusion of fluorine adatoms on doped graphene},}\ }\href {\doibase
  10.1063/1.4896511} {\bibfield  {journal} {\bibinfo  {journal} {Appl. Phys.
  Lett.}\ }\textbf {\bibinfo {volume} {105}},\ \bibinfo {pages} {121606}
  (\bibinfo {year} {2014})}\BibitemShut {NoStop}%
\bibitem [{\citenamefont {{Irmer}}\ \emph {et~al.}(2014)\citenamefont
  {{Irmer}}, \citenamefont {{Frank}}, \citenamefont {{Putz}}, \citenamefont
  {{Gmitra}}, \citenamefont {{Kochan}},\ and\ \citenamefont
  {{Fabian}}}]{Irmer2014}%
  \BibitemOpen
  \bibfield  {author} {\bibinfo {author} {\bibfnamefont {S.}~\bibnamefont
  {{Irmer}}}, \bibinfo {author} {\bibfnamefont {T.}~\bibnamefont {{Frank}}},
  \bibinfo {author} {\bibfnamefont {S.}~\bibnamefont {{Putz}}}, \bibinfo
  {author} {\bibfnamefont {M.}~\bibnamefont {{Gmitra}}}, \bibinfo {author}
  {\bibfnamefont {D.}~\bibnamefont {{Kochan}}}, \ and\ \bibinfo {author}
  {\bibfnamefont {J.}~\bibnamefont {{Fabian}}},\ }\bibfield  {title} {\enquote
  {\bibinfo {title} {Spin-orbit coupling in fluorinated graphene},}\
  }\href@noop {} {\bibfield  {journal} {\bibinfo  {journal} {ArXiv e-prints}\ }
  (\bibinfo {year} {2014})},\ \Eprint {http://arxiv.org/abs/1411.0016}
  {arXiv:1411.0016 [cond-mat.mes-hall]} \BibitemShut {NoStop}%
\bibitem [{\citenamefont {Giannozzi}\ \emph {et~al.}(2009)\citenamefont
  {Giannozzi}, \citenamefont {Baroni}, \citenamefont {Bonini}, \citenamefont
  {Calandra}, \citenamefont {Car}, \citenamefont {Cavazzoni}, \citenamefont
  {Ceresoli}, \citenamefont {Chiarotti}, \citenamefont {Cococcioni},
  \citenamefont {Dabo}, \citenamefont {Corso}, \citenamefont {de~Gironcoli},
  \citenamefont {Fabris}, \citenamefont {Fratesi}, \citenamefont {Gebauer},
  \citenamefont {Gerstmann}, \citenamefont {Gougoussis}, \citenamefont
  {Kokalj}, \citenamefont {Lazzeri}, \citenamefont {Martin-Samos},
  \citenamefont {Marzari}, \citenamefont {Mauri}, \citenamefont {Mazzarello},
  \citenamefont {Paolini}, \citenamefont {Pasquarello}, \citenamefont
  {Paulatto}, \citenamefont {Sbraccia}, \citenamefont {Scandolo}, \citenamefont
  {Sclauzero}, \citenamefont {Seitsonen}, \citenamefont {Smogunov},
  \citenamefont {Umari},\ and\ \citenamefont {Wentzcovitch}}]{QE}%
  \BibitemOpen
  \bibfield  {author} {\bibinfo {author} {\bibfnamefont {P.}~\bibnamefont
  {Giannozzi}}, \bibinfo {author} {\bibfnamefont {S.}~\bibnamefont {Baroni}},
  \bibinfo {author} {\bibfnamefont {N.}~\bibnamefont {Bonini}}, \bibinfo
  {author} {\bibfnamefont {M.}~\bibnamefont {Calandra}}, \bibinfo {author}
  {\bibfnamefont {R.}~\bibnamefont {Car}}, \bibinfo {author} {\bibfnamefont
  {C.}~\bibnamefont {Cavazzoni}}, \bibinfo {author} {\bibfnamefont
  {D.}~\bibnamefont {Ceresoli}}, \bibinfo {author} {\bibfnamefont {G.~L.}\
  \bibnamefont {Chiarotti}}, \bibinfo {author} {\bibfnamefont {M.}~\bibnamefont
  {Cococcioni}}, \bibinfo {author} {\bibfnamefont {I.}~\bibnamefont {Dabo}},
  \bibinfo {author} {\bibfnamefont {A.~D.}\ \bibnamefont {Corso}}, \bibinfo
  {author} {\bibfnamefont {S.}~\bibnamefont {de~Gironcoli}}, \bibinfo {author}
  {\bibfnamefont {S.}~\bibnamefont {Fabris}}, \bibinfo {author} {\bibfnamefont
  {G.}~\bibnamefont {Fratesi}}, \bibinfo {author} {\bibfnamefont
  {R.}~\bibnamefont {Gebauer}}, \bibinfo {author} {\bibfnamefont
  {U.}~\bibnamefont {Gerstmann}}, \bibinfo {author} {\bibfnamefont
  {C.}~\bibnamefont {Gougoussis}}, \bibinfo {author} {\bibfnamefont
  {A.}~\bibnamefont {Kokalj}}, \bibinfo {author} {\bibfnamefont
  {M.}~\bibnamefont {Lazzeri}}, \bibinfo {author} {\bibfnamefont
  {L.}~\bibnamefont {Martin-Samos}}, \bibinfo {author} {\bibfnamefont
  {N.}~\bibnamefont {Marzari}}, \bibinfo {author} {\bibfnamefont
  {F.}~\bibnamefont {Mauri}}, \bibinfo {author} {\bibfnamefont
  {R.}~\bibnamefont {Mazzarello}}, \bibinfo {author} {\bibfnamefont
  {S.}~\bibnamefont {Paolini}}, \bibinfo {author} {\bibfnamefont
  {A.}~\bibnamefont {Pasquarello}}, \bibinfo {author} {\bibfnamefont
  {L.}~\bibnamefont {Paulatto}}, \bibinfo {author} {\bibfnamefont
  {C.}~\bibnamefont {Sbraccia}}, \bibinfo {author} {\bibfnamefont
  {S.}~\bibnamefont {Scandolo}}, \bibinfo {author} {\bibfnamefont
  {G.}~\bibnamefont {Sclauzero}}, \bibinfo {author} {\bibfnamefont {A.~P.}\
  \bibnamefont {Seitsonen}}, \bibinfo {author} {\bibfnamefont {A.}~\bibnamefont
  {Smogunov}}, \bibinfo {author} {\bibfnamefont {P.}~\bibnamefont {Umari}}, \
  and\ \bibinfo {author} {\bibfnamefont {R.~M.}\ \bibnamefont {Wentzcovitch}},\
  }\bibfield  {title} {\enquote {\bibinfo {title} {Quantum espresso: a modular
  and open-source software project for quantum simulations of materials},}\
  }\href {http://www.quantum-espresso.org} {\bibfield  {journal} {\bibinfo
  {journal} {J. Phys.: Condens. Matter}\ }\textbf {\bibinfo {volume} {21}},\
  \bibinfo {pages} {395502} (\bibinfo {year} {2009})}\BibitemShut {NoStop}%
\bibitem [{\citenamefont {Perdew}\ \emph {et~al.}(1996)\citenamefont {Perdew},
  \citenamefont {Burke},\ and\ \citenamefont {Ernzerhof}}]{pbe}%
  \BibitemOpen
  \bibfield  {author} {\bibinfo {author} {\bibfnamefont {J.~P.}\ \bibnamefont
  {Perdew}}, \bibinfo {author} {\bibfnamefont {K.}~\bibnamefont {Burke}}, \
  and\ \bibinfo {author} {\bibfnamefont {M.}~\bibnamefont {Ernzerhof}},\
  }\bibfield  {title} {\enquote {\bibinfo {title} {Generalized gradient
  approximation made simple},}\ }\href@noop {} {\bibfield  {journal} {\bibinfo
  {journal} {Phys. Rev. Lett.}\ }\textbf {\bibinfo {volume} {77}},\ \bibinfo
  {pages} {3865} (\bibinfo {year} {1996})}\BibitemShut {NoStop}%
\bibitem [{\citenamefont {Vanderbilt}(1990)}]{Vanderbilt1990}%
  \BibitemOpen
  \bibfield  {author} {\bibinfo {author} {\bibfnamefont {D.}~\bibnamefont
  {Vanderbilt}},\ }\bibfield  {title} {\enquote {\bibinfo {title} {Soft
  self-consistent pseudopotentials in a generalized eigenvalue formalism},}\
  }\href {\doibase 10.1103/PhysRevB.41.7892} {\bibfield  {journal} {\bibinfo
  {journal} {Phys. Rev. B}\ }\textbf {\bibinfo {volume} {41}},\ \bibinfo
  {pages} {7892} (\bibinfo {year} {1990})}\BibitemShut {NoStop}%
\bibitem [{\citenamefont {Neugebauer}\ and\ \citenamefont
  {Scheffler}(1992)}]{Neugebauer1992}%
  \BibitemOpen
  \bibfield  {author} {\bibinfo {author} {\bibfnamefont {J.}~\bibnamefont
  {Neugebauer}}\ and\ \bibinfo {author} {\bibfnamefont {M.}~\bibnamefont
  {Scheffler}},\ }\bibfield  {title} {\enquote {\bibinfo {title}
  {Adsorbate-substrate and adsorbate-adsorbate interactions of na and k
  adlayers on al(111)},}\ }\href@noop {} {\bibfield  {journal} {\bibinfo
  {journal} {Phys. Rev. B}\ }\textbf {\bibinfo {volume} {46}},\ \bibinfo
  {pages} {16067} (\bibinfo {year} {1992})}\BibitemShut {NoStop}%
\bibitem [{\citenamefont {Pereira}\ \emph {et~al.}(2008)\citenamefont
  {Pereira}, \citenamefont {Lopes~dos Santos},\ and\ \citenamefont
  {Castro~Neto}}]{Pereira2008}%
  \BibitemOpen
  \bibfield  {author} {\bibinfo {author} {\bibfnamefont {V.~M.}\ \bibnamefont
  {Pereira}}, \bibinfo {author} {\bibfnamefont {J.}~\bibnamefont {Lopes~dos
  Santos}}, \ and\ \bibinfo {author} {\bibfnamefont {A.~H.}\ \bibnamefont
  {Castro~Neto}},\ }\bibfield  {title} {\enquote {\bibinfo {title} {Modeling
  disorder in graphene},}\ }\href@noop {} {\bibfield  {journal} {\bibinfo
  {journal} {Phys. Rev. B}\ }\textbf {\bibinfo {volume} {77}},\ \bibinfo
  {pages} {115109} (\bibinfo {year} {2008})}\BibitemShut {NoStop}%
\bibitem [{\citenamefont {Cheianov}\ \emph {et~al.}(2010)\citenamefont
  {Cheianov}, \citenamefont {Sylju{\aa}sen}, \citenamefont {Altshuler},\ and\
  \citenamefont {Fal'ko}}]{Cheianov2010}%
  \BibitemOpen
  \bibfield  {author} {\bibinfo {author} {\bibfnamefont {V.~V.}\ \bibnamefont
  {Cheianov}}, \bibinfo {author} {\bibfnamefont {O.}~\bibnamefont
  {Sylju{\aa}sen}}, \bibinfo {author} {\bibfnamefont {B.~L.}\ \bibnamefont
  {Altshuler}}, \ and\ \bibinfo {author} {\bibfnamefont {V.~I.}\ \bibnamefont
  {Fal'ko}},\ }\bibfield  {title} {\enquote {\bibinfo {title} {Sublattice
  ordering in a dilute ensemble of monovalent adatoms on graphene},}\ }\href
  {http://stacks.iop.org/0295-5075/89/i=5/a=56003} {\bibfield  {journal}
  {\bibinfo  {journal} {EPL (Europhysics Letters)}\ }\textbf {\bibinfo {volume}
  {89}},\ \bibinfo {pages} {56003} (\bibinfo {year} {2010})}\BibitemShut
  {NoStop}%
\bibitem [{\citenamefont {Mostofi}\ \emph {et~al.}(2008)\citenamefont
  {Mostofi}, \citenamefont {Yates}, \citenamefont {Lee}, \citenamefont {Souza},
  \citenamefont {Vanderbilt},\ and\ \citenamefont {Marzari}}]{Mostofi2008}%
  \BibitemOpen
  \bibfield  {author} {\bibinfo {author} {\bibfnamefont {A.~A.}\ \bibnamefont
  {Mostofi}}, \bibinfo {author} {\bibfnamefont {J.~R.}\ \bibnamefont {Yates}},
  \bibinfo {author} {\bibfnamefont {Y.-S.}\ \bibnamefont {Lee}}, \bibinfo
  {author} {\bibfnamefont {I.}~\bibnamefont {Souza}}, \bibinfo {author}
  {\bibfnamefont {D.}~\bibnamefont {Vanderbilt}}, \ and\ \bibinfo {author}
  {\bibfnamefont {N.}~\bibnamefont {Marzari}},\ }\bibfield  {title} {\enquote
  {\bibinfo {title} {wannier90: A tool for obtaining maximally-localised
  wannier functions},}\ }\href {\doibase
  http://dx.doi.org/10.1016/j.cpc.2007.11.016} {\bibfield  {journal} {\bibinfo
  {journal} {Computer Physics Communications}\ }\textbf {\bibinfo {volume}
  {178}},\ \bibinfo {pages} {685 } (\bibinfo {year} {2008})}\BibitemShut
  {NoStop}%
\bibitem [{\citenamefont {Marzari}\ \emph {et~al.}(2012)\citenamefont
  {Marzari}, \citenamefont {Mostofi}, \citenamefont {Yates}, \citenamefont
  {Souza},\ and\ \citenamefont {Vanderbilt}}]{Marzari2012}%
  \BibitemOpen
  \bibfield  {author} {\bibinfo {author} {\bibfnamefont {N.}~\bibnamefont
  {Marzari}}, \bibinfo {author} {\bibfnamefont {A.~A.}\ \bibnamefont
  {Mostofi}}, \bibinfo {author} {\bibfnamefont {J.~R.}\ \bibnamefont {Yates}},
  \bibinfo {author} {\bibfnamefont {I.}~\bibnamefont {Souza}}, \ and\ \bibinfo
  {author} {\bibfnamefont {D.}~\bibnamefont {Vanderbilt}},\ }\bibfield  {title}
  {\enquote {\bibinfo {title} {Maximally localized wannier functions: Theory
  and applications},}\ }\href {\doibase 10.1103/RevModPhys.84.1419} {\bibfield
  {journal} {\bibinfo  {journal} {Rev. Mod. Phys.}\ }\textbf {\bibinfo {volume}
  {84}},\ \bibinfo {pages} {1419} (\bibinfo {year} {2012})}\BibitemShut
  {NoStop}%
\bibitem [{\citenamefont {Radford}\ \emph {et~al.}(1961)\citenamefont
  {Radford}, \citenamefont {Hughes},\ and\ \citenamefont
  {Beltran-Lopez}}]{Radford1961}%
  \BibitemOpen
  \bibfield  {author} {\bibinfo {author} {\bibfnamefont {H.~E.}\ \bibnamefont
  {Radford}}, \bibinfo {author} {\bibfnamefont {V.~W.}\ \bibnamefont {Hughes}},
  \ and\ \bibinfo {author} {\bibfnamefont {V.}~\bibnamefont {Beltran-Lopez}},\
  }\bibfield  {title} {\enquote {\bibinfo {title} {Microwave zeeman spectrum of
  atomic fluorine},}\ }\href {\doibase 10.1103/PhysRev.123.153} {\bibfield
  {journal} {\bibinfo  {journal} {Phys. Rev.}\ }\textbf {\bibinfo {volume}
  {123}},\ \bibinfo {pages} {153} (\bibinfo {year} {1961})}\BibitemShut
  {NoStop}%
\bibitem [{\citenamefont {Petersen}\ and\ \citenamefont
  {Hedeg{\aa}rd}(2000)}]{Petersen2000}%
  \BibitemOpen
  \bibfield  {author} {\bibinfo {author} {\bibfnamefont {L.}~\bibnamefont
  {Petersen}}\ and\ \bibinfo {author} {\bibfnamefont {P.}~\bibnamefont
  {Hedeg{\aa}rd}},\ }\bibfield  {title} {\enquote {\bibinfo {title} {A simple
  tight-binding model of spin{\textendash}orbit splitting of sp-derived surface
  states},}\ }\href@noop {} {\bibfield  {journal} {\bibinfo  {journal} {Surf.
  Sci.}\ }\textbf {\bibinfo {volume} {459}},\ \bibinfo {pages} {49} (\bibinfo
  {year} {2000})}\BibitemShut {NoStop}%
\bibitem [{\citenamefont {Ast}\ and\ \citenamefont {Gierz}(2012)}]{Ast2012}%
  \BibitemOpen
  \bibfield  {author} {\bibinfo {author} {\bibfnamefont {C.~R.}\ \bibnamefont
  {Ast}}\ and\ \bibinfo {author} {\bibfnamefont {I.}~\bibnamefont {Gierz}},\
  }\bibfield  {title} {\enquote {\bibinfo {title} {$sp$-band tight-binding
  model for the bychkov-rashba effect in a two-dimensional electron system
  including nearest-neighbor contributions from an electric field},}\ }\href
  {\doibase 10.1103/PhysRevB.86.085105} {\bibfield  {journal} {\bibinfo
  {journal} {Phys. Rev. B}\ }\textbf {\bibinfo {volume} {86}},\ \bibinfo
  {pages} {085105} (\bibinfo {year} {2012})}\BibitemShut {NoStop}%
\end{thebibliography}

%

\end{document}